\documentclass[a4paper,11pt]{article}
\pdfoutput=1 

\usepackage{jheppub} 

\usepackage{graphicx}
\usepackage[figurename=Fig.]{caption}
\usepackage[tablename=Tab.]{caption}
\usepackage{color}
\usepackage{float}
\usepackage{amssymb}
\usepackage{amsmath}
\usepackage{mathrsfs}
\usepackage{textcomp}
\usepackage{bbm}
\usepackage{cases}
\usepackage{makecell}
\usepackage{pdflscape}
\usepackage{subfigure}
\usepackage{pdfpages}
\usepackage{makeidx}
\usepackage{bm}
\usepackage{xstring}
\usepackage{lipsum}
\usepackage{slashed}
\usepackage{multicol}
\usepackage{setspace}
\usepackage{booktabs}
\usepackage[table]{xcolor}
\usepackage{multirow}
\usepackage{braket}
\usepackage[separate-uncertainty]{siunitx}
\usepackage[displaymath, mathlines]{lineno}

\newcommand{\eref}[1]{Eq.\,(\ref{#1})}

\newcommand{\vo}{\vec{o}\@ifnextchar{^}{\,}{}}
\def\up{\mathrm}
\def\d{\up{d}}

\renewcommand{\thetable}{\Roman{table}}
\newcommand{\vabs}[1]{|\vec #1\,|}

\graphicspath{{PDF/}}

\def\slash#1{\setbox0=\hbox{$#1$}           
   \dimen0=\wd0                                 
   \setbox1=\hbox{/} \dimen1=\wd1               
   \ifdim\dimen0>\dimen1                        
      \rlap{\hbox to \dimen0{\hfil/\hfil}}      
      #1                                        
   \else                                        
      \rlap{\hbox to \dimen1{\hfil$#1$\hfil}}   
      /                                         
   \fi}                                         %
\def\sl#1{\setbox0=\hbox{#1} 
  \dimen0=\wd0 
  \rlap{\hbox to \dimen0{\hss/\hss}}%
  #1} 

\title{Mass spectra and wave functions of $T_{QQ\bar Q\bar Q}$ tetraquarks}

\author[1]{Qiang Li}
\author[2,3,4]{Chao-Hsi Chang}
\author[5]{Guo-Li Wang}
\author[6]{Tianhong Wang}

\affiliation[1]{School of Physical Science and Technology, Northwestern Polytechnical University, Xi'an 710072, China}
\affiliation[2]{Institute of Theoretical Physics, Chinese Academy of Sciences, Beijing 100190, China}
\affiliation[3]{School of Physical Sciences, University of Chinese Academy of Sciences, 19A Yuquan Road, Beijing 100049, China}
\affiliation[4]{CCAST (World Laboratory), P.O. Box 8730, Beijing 100190, China}
\affiliation[5]{Department of Physics, Hebei University, Baoding 071002, China}
\affiliation[6]{School of Physics, Harbin Institute of Technology, Harbin 150001, China}

\emailAdd{liruo@nwpu.edu.cn}
\emailAdd{zhangzx@itp.ac.cn}
\emailAdd{gl\_wang@hit.edu.cn}
\emailAdd{thwang@hit.edu.cn}

\abstract{
The compact tetraquark states with fully heavy quark contents $QQ\bar Q\bar Q$ are studied as the bound states of the diquark-antidiquark within the Bethe-Salpeter framework. The (anti)diquark masses and form factors used are the same as we calculated the doubly heavy baryons in a previous work. Under the instantaneous approximation, the three-dimensional (Bethe-)Salpeter equation of the tetraquarks is derived and solved numerically to obtain the corresponding mass spectra and wave functions of the tetraquarks with $J^{PC}=0^{++}$, $1^{+-}$, and $2^{++}$.  Our results show that the three ground states of $cc\bar c\bar c$ locate in the mass range of $6.4\sim6.5\,\si{GeV}$, and the $bb\bar b\bar b$ states in mass range of $19.2\sim19.3\,\si{GeV}$. The obtained relativistic wave functions naturally include the mixing effects from the possible $D$\,(or $G$) partial waves, and then can be further used to do  precise calculations of the tetraquark decays. Based on the obtained results, the LHCb's observation $X(6900)$ is less likely to be the ground states of compact $cc\bar c\bar c$ tetraquarks but might be the first or second excited states. In addition, a widely used propagator-like form factor is also investigated and discussed.
}

\begin{document} 
\maketitle
\flushbottom

\section{Introduction}
The quantum chromodynamics and the quark model allows not only the well-known traditional hadrons, such as the $q\bar q$-type mesons and $qqq$-type baryons, but also the exotic tetraquark states and the pentaquark baryons\,\cite{GellMann1964,Zweig1964}. About 50 years after the predictions of these exotic states, the LHCb Collaboration first detected the pentaquark baryons\,\cite{LHCb2015-Pc,LHCb2019-Pc}, and then in 2020 reported a new narrow structure labeled as $X(6900)$ which covers the predicted masses of states
composed of four charm quarks\,\cite{LHCb2020-X6900}. 

Although not the first hint of the tetraquark states, $X(6900)$ causes great attention in hadron physics for its fully heavy quark contents. Inspired by this observation,  the four-charm states around 6.9 GeV have been investigated in several models or approaches\,\cite{ZhaoJX2020,Gordillo2020,Mutuk2021,Giron2020,Lundhammar2020,DengCR2021,WengXZ2021,WangZG2020,LuQF2020,Faustov2020,Bedolla2020,KeHW2021,ZhuRL2021,ChaoKT2020,Karliner2020}, 
{\color{black}
such as, solving the two- or many-body time-independent Schr\"odinger equation\,\cite{ZhaoJX2020,Gordillo2020,Mutuk2021,Giron2020,Lundhammar2020}, the chromomagnetic interaction models\,\cite{DengCR2021,WengXZ2021},  the QCD sum rules combined with the Regge trajectories\,\cite{WangZG2020}, the extended Godfrey and Isgur\,(GI) quark model\,\cite{LuQF2020}, the relativistic quark model based on the quasipotential approach or effective Hamiltonian\,\cite{Faustov2020,Bedolla2020}, and also the Bethe-Salpeter framework with different interaction kernels and approximation methods\,\cite{KeHW2021,ZhuRL2021}.  Also notice by using the heavy diquark limit $M_i\to\infty$,  Ref.\,\cite{ZhuRL2021} is in fact dealing with a  Schr\"odinger equation combined with the Regge trajectories to obtain the mass spectra. Notice most of these previous studies are based on the nonrelativistic Schr\"odinger equation or effective Hamiltonian methods while the relativistic effects and the possible $S$-$D$ or $D$-$G$ mixing effects are not included properly}. 
In these previous studies, the ground states of the $cc\bar c\bar c$ structures are predicted to be $6.1\sim 6.5$ GeV, and the $X(6900)$ are usually tentatively identified as the radial or orbital excited states of tetraquark $cc\bar c\bar c$ in these recent studies, or interpreted as the coupled-channel or rescattering effects of two (or more) charmonia\,\cite{GuoZH2021,DongXK2021,WangJZ2021}. Other interpretations, such as the gluonic tetracharm\,\cite{WanBD2020} or light Higgs-like boson\,\cite{ZhuJW2020} are also proposed. So far, there is still no strong evidences whether these observed exotic hadrons are the genuine multiquark bound states or just the loosely bound molecules of the traditional mesons and baryons.

The fully heavy tetraquark states have several advantages in both experimental and theoretical researches. On the one hand, the mass of fully heavy tetraquarks locates far away from those of the traditional mesons and doubly heavy tetraquark $Q\bar Q q\bar q$ and they can be clearly identified from the the known hadron spectra. On the other hand, the molecular states of two charmonia can not be bounded by the light boson exchange which makes the molecule configuration much more difficult to produce such states. Hence, the exotic states consisting of four heavy quarks are more likely to be the real compact tetraquarks.   

In this work, we will try to deal with the mass spectra and wave functions of the fully heavy tetraquark states $T_{QQ\bar Q\bar Q}$, namely, $cc\bar c\bar c$ or $bb\bar b\bar b$, within the framework of the Bethe-Salpeter equation. The fully heavy tetraquarks are assumed to be formed by the diquark $QQ$ and the antidiquark $\bar Q\bar Q$. The diquark $QQ$ is further assumed to be in the color $\bar 3$ configuration in order to produce the attractive force, and then similarly, the antidiquark $\bar Q\bar Q$ is in the color $3$ state. This compact color-$\bar 3$ diquark picture has already been used in a previous work to study the doubly heavy baryons\,\cite{LiQ2020}, where the obtained mass of $\Xi_{cc}^{++}$ is just 20 MeV lower than the experimental measurements, and other predictions are also consistent with the recent theoretical researches especially the Lattice QCD results\,\cite{Brown2014}. We will use the previously calculated mass spectra and form factors of the $J^P=1^+$ $cc$ and $bb$ diquarks in this work. {\color{black} In addition, a propagator-like form factor is investigated and the corresponding cutoff dependence is also studied.} The color $\bar 3$ $QQ$ diquark and the color $3$ $\bar Q\bar Q$ antidiquark can finally form a compact tetraquark in color singlet by the one-gluon-exchange interaction.  Based on the above analysis, the four-body tetraquark problem can be first reduced into two two-body bound problems of fermion\,(antifermon) system, which can be solved by calculating the original  Bethe-Salpeter equation (BSE)\cite{SB1951,Salpeter1952}.  Then we need to deal with a two-body problem of boson system which will be the main focus of this work. 

The Bethe-Salpeter framework has great advantages in dealing with the two-body  bound states for the relativistic interaction kernel and corresponding wave functions. The constructed relativistic wave functions are based on the good quantum number $J^{P(C)}$ rather than the nonrelativistic characteristics spin and orbital angular momentum $^{2S+1}L_J$. The BSE framework has been successfully used in the mass spectra of mesons\,\cite{Chang2005A,Chang2010,LiQ2019A}, traditional baryons and pentaquarks\,\cite{LiQ2020,XuH2020}, hadronic transitions, electro-weak decays, and etc\,\cite{Chang2005,WangZ2012A,WangT2013,WangT2013A,LiQ2016,LiQ2017,LiQ2017A}. In this work we would try to push the BSE framework further to study the fully heavy $QQ\bar Q\bar Q$ system and develop a precise and systematic approach to describe the compact tetraquark states.

This manuscript is organized as: in section \ref{Sec-2} the (Bethe-)Salpeter equation is derived in the perspective of the tetraquarks taken as the axial-vector diquark-antidiquark bound states; in section \ref{Sec-3} the Salpeter wave functions of the tetraquarks with $J^{PC}=0^{++}$, $1^{+-}$, and $2^{++}$ are constructed; then in section \ref{Sec-4} the obtained mass spectra and wave functions are given and discussed; finally a brief summary is presented.

\section{Tetraquarks as the bound states of diquark and antidiquark}\label{Sec-2}

Considering the exclusive principle, the diquark\,(antidiquark), consisting of two charm quarks\,(antiquarks) in the color $\bar 3$\,(3) configuration, could only be in the $J^P=1^+$ spin configuration in orbital ground states, since the flavor wave function is naturally symmetric and spatial wave function is also symmetric in ground states. In current work, we do not consider the excitation of the diquarks or antidiquarks. Now we try to deal with the BSE of the tetraquark states consisting of a $1^+$ diquark and a $1^+$ antidiquark, which could form three ground states with spin-parity configuration $J^{PC}=0^{++}$, $1^{+-}$, and $2^{++}$ respectively.

\subsection{Bethe-Salpeter equation of two vector bosons}
\begin{figure}[htpb]
\centering
\includegraphics[width=0.7\textwidth]{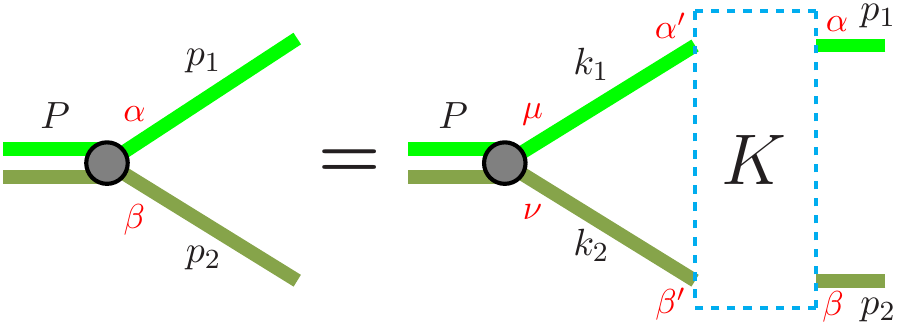} \label{Fig-BSE-1+1}
\caption{Bethe-Salpeter equation of the tetraquark states in the diquark-antidiquark picture.  The Greeks~(red) are used for the Lorentz indices. $P,~p_1(k_1),~p_2(k_2)$ denote the momenta of the tetraquark state, constituent diquark, and the constituent antidiquark respectively; $K_{\alpha\alpha';\beta'\beta}$ denotes the effective interaction kernel between the diquark and the antidiquark based on the one-gluon exchange.}\label{Fig-BS-T}
\end{figure}

The Bethe-Salpeter equations of the bound states consisting of two vector (or axialvactor) constituents are schematically depicted in \autoref{Fig-BS-T}. The corresponding BSE are expressed as the four-dimensional integral of the inner relative momentum $k$, 
\begin{gather}
\Gamma^{\alpha\beta}(P,q,\xi) =\int \frac{\up d^4 k}{(2\pi)^4} (-i)K^{\alpha\alpha';\beta'\beta}(P,k,q)[D_{\alpha'\mu}(k_2) \Gamma^{\mu\nu}(P,k,\xi) D_{\nu\beta'}(k_1)], \label{E-BSE-T} 
\end{gather}
where $\Gamma^{\alpha\beta}(P,q,\xi)$ denotes the vertex of the two axialvector constituents; symbol $P$ is used to denote the momentum of the tetraquark state, and we have $P^2=M^2$ with $M$ representing the tetraquark mass; symbol $\xi$ indicates the polarization state, with $\xi=\pm2, 0$ for the $J^P=2^+$ tetraquark states, $\xi=\pm1,0$ for the $1^+$ ones, and $\xi=0$ for the scalar ones; $K^{\alpha\alpha';\beta'\beta}(P,k,q)$ is the interaction kernel of the diquark pair based on the one-gluon exchange. The constituent mass of the diquark\,(antidiquark) is represented by $M_{1(2)} $. The constituent masses of the $J^P=1^+$ $cc$ and $bb$ diquarks are obtained by solving the corresponding BSE\,\cite{LiQ2020}. The inner relative momenta $q$  and $k$ are defined as $q=\alpha_{2}p_1-\alpha_{1}p_2$ and $k=\alpha_{2}k_1-\alpha_{1} k_2$ respectively, with $\alpha_{i}\equiv \frac{M_{i}}{M_{1}+M_2}$.  The effective propagator of the $1^+$ diquark reads,
\begin{gather} 
D^{\alpha\beta}(k_1)  =D(k_1)d^{\alpha\beta}(k_{1\perp}),~~~d^{\alpha\beta}(k_{1\perp})= -g^{\alpha\beta}+\frac{k_{1\perp}^\alpha k_{1\perp}^\beta}{M_{1}^2},
\end{gather}
where $k_{i\perp}\equiv k_i- \frac{k_i\cdot P}{M}$, and $D(k_1)=i\frac{1}{k_1^2-M_{1}^2+i\epsilon}$ is the usual scalar propagator.
As usual, the Bethe-Salpeter wave function describing the tetraquark states can be defined as
\begin{align}
T_{\alpha\beta}(P,q,\xi)=D_{\alpha\mu}(p_2) \Gamma^{\mu\nu}(P,q,\xi) D_{\beta\nu}(p_1). \label{E-BS-wave}
\end{align}
The symbols $P$ and $\xi$ in the BS wave function $T_{\alpha\beta}(P,q,\xi)$ and vertex $\Gamma^{\mu\nu}(P,q,\xi)$ will be omitted unless it is necessary to specify them. By using the definition of BS wave function, the BSE in \eref{E-BSE-T} can also be rewritten as the integral of $T_{\alpha\beta}$,
\begin{gather}
D^{-1}_{\alpha\alpha'} (p_1)  T_{\alpha'\beta'}(q) D^{-1}_{\beta'\beta}(p_2)=  \int \frac{\d^4k}{(2\pi)^4}
(-i) K^{\alpha\alpha';\beta'\beta}(P,k,q) T_{\alpha'\beta'}(k),
\end{gather} 
where the inverse of the vector propagator reads,
\begin{gather} 
D^{-1}_{\alpha\beta}(p_i)=\vartheta_{\alpha\beta}(p_i)D^{-1}(p_i),~~~\vartheta^{\alpha\beta}(p_i)=-g^{\alpha\beta}- \frac{p^\alpha_{i\perp} p^\beta_{i\perp}}{w_{i}^2}.
\end{gather}
It can be easily checked that $\vartheta^{\alpha\beta}(p_i)$ fulfills the condition $\vartheta^{\alpha\beta}(p_i)d_{\beta\gamma}(p_i)=\delta^{\alpha}_{\gamma}$, where $w_i\equiv (M_i^2-p_{i\perp}^2)^{\frac12}$ denotes the kinetic energy of the $i$th constituent.

\subsection{Diquark form factors and tetraquark interaction kernel}

For the tetraquark states consisting of the $1^+$ diquark and antidiquark constituents, each vector constituent has internal structure and usually can not be regarded as the pointlike particle.  The diquark (antidiquark) is usually described by the corresponding form factors.  Generally speaking, the form factors have great effects on the energy splittings of the tetraquark bound states. 
Then the potential $V(s)$ between the two constituents will be smeared by the two form factors. With these two form factors, the interaction kernels of tetraquarks can be expressed as,
\begin{gather} \label{E-Kernel}
K^{\alpha\alpha';\beta'\beta}(P,k,q) =f_\up{1V}f_\up{2V} g^{\alpha\alpha'} g^{\beta'\beta}(k_1+p_{1})^\nu  (k_2+p_2)_\nu V(s),
\end{gather}
where $f_\up{1V}$ and $f_\up{2V}$ denote the vector form factor of the diquark and antidiquark, and in the fully heavy $cc\bar c\bar c$ or $bb\bar b\bar b$ system, we have $f_{1\up{V}}=f_\up{2V}=f_\up{V}$; $V(s)$ denotes the one-gluon exchange potential with $s\equiv (k-q)$ being the momentum of the exchanged gluon. Compared with the case in baryon problem\,\cite{LiQ2020}, there are two form factors to describe the nonpointlike structures, which make it much more complicated to deal with the tetraquark system.

Under the instantaneous approximation, the one-gluon-exchange potential is assumed to be static, namely, $V(s)\sim V(s_\perp)$ with $s_\perp=s-\frac{P\cdot s}{M}$, which reads in the Coulomb gauge as\,\cite{Chao1992,DingYB1993,DingYB1995,Kim2004}
\begin{equation}
V(\vec s\,)=V_\up{Coul} + V_{\up{Conf}}= -\frac{4}{3} \frac{4\pi \alpha_s(\vec s\,)}{\vec s\,^2+a_1^2}+\left[(2\pi)^3 \delta^3(\vec s\,)\left( \frac{\lambda}{a_2}+V_0 \right)- \frac{8\pi \lambda}{(\vec s\,^2+a_2^2)^2} \right],
\end{equation}
where $\frac{4}{3}$ is the color factor in the color singlet; $a_{1(2)}$ is introduced to avoid the divergence in small momentum transfer zone; the potential $V_\up{Conf}(\vec s\,)$ describing the confinement effects is introduced phenomenologically, which is characterized by the the string constant $\lambda$ and the factor $a_2$. The potential used here is based on the famous Cornell potential\,\cite{Eichten1978,Eichten1980}, which behaves as the one-gluon exchange Coulomb-type potential at short distance and a linear growth confinement one at long distance, and then modified as the aforementioned one to incorporate the color screening effects\,\cite{Laermann1986,Born1989} in the linear confinement potential. $V_0$ is a constant fixed by fitting to the meson data. The strong coupling constant $\alpha_s$ has the following form, 
\[\alpha_s(\vec s\,)=\frac{12\pi}{(33-2N_f)}\frac{1}{\ln\left(a+ {\vec s\,^2}/{\Lambda^2_{\up{QCD}}}\right)},\]
where $\Lambda_\up{QCD}$ is the scale of the strong interaction; $N_f$ is the active flavor number which is 3 for the $cc$ interaction while 4 for the $bb$ interaction; $a=e$ is a regulator constant. For later convenience,  we further split $V(\vec s\,)$ into two parts as
\begin{equation}
V(\vec s\,)=(2\pi)^3\delta^3 (\vec s\,) V_1+V_2 (\vec s\,).
\end{equation}
Namely, all the dependence on $\vec s$ is incorporated into $V_{2}(\vec s\,)$, while $V_{1}$ is just the constant item.

Also since the spatial parts are suppressed by a factor of $(\frac{v}{c})^2$, we only consider the dominate time component ($\nu=0$) in the one-gluon-exchange Lorentz  structure, namely, 
\begin{gather}
 (k_1+p_{1})^\nu  (k_2+p_2)_\nu \simeq \varrho_1 \varrho_2,
 \end{gather} 
where we used the abbreviations $\varrho_i \equiv 2(\alpha_i M+ q_P)$ with definition $q_P\equiv \frac{P\cdot q}{M}$.
Using the notations and approximations introduced above, the tetraquark interaction kernel behaves as
\begin{align} \label{E-Kernel-3D}
K^{\alpha\alpha';\beta'\beta}(P,k,q) \simeq \varrho_1 \varrho_2 g^{\alpha\alpha'} g^{\beta'\beta} \varkappa\left(s_\perp\right),
\end{align}
where $\varkappa \equiv f_\up{V}^2 V(\vec s\,)$; the factor $\varrho_1 \varrho_2$ has been stripped off for later convenience. Notice the kernel $\varkappa(s_\perp)$ here has no dependence on the time component of momentum transfer $s$.

\begin{figure}[htbp]
\vspace{0.5em}
\centering
\subfigure[Method I]{\includegraphics[width=0.48\textwidth]{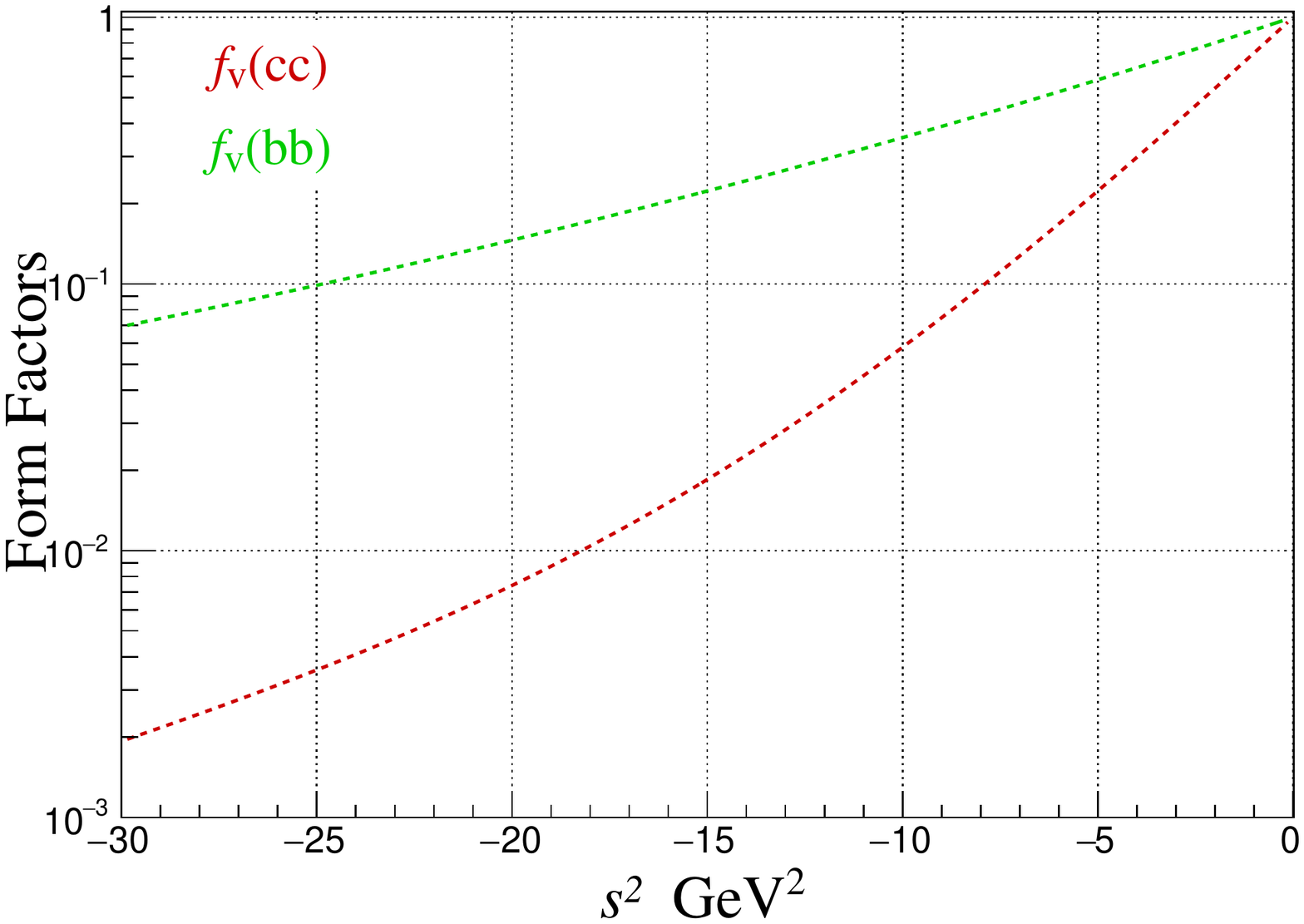} \label{Fig-form-I}}
\subfigure[Method II]{\includegraphics[width=0.48\textwidth]{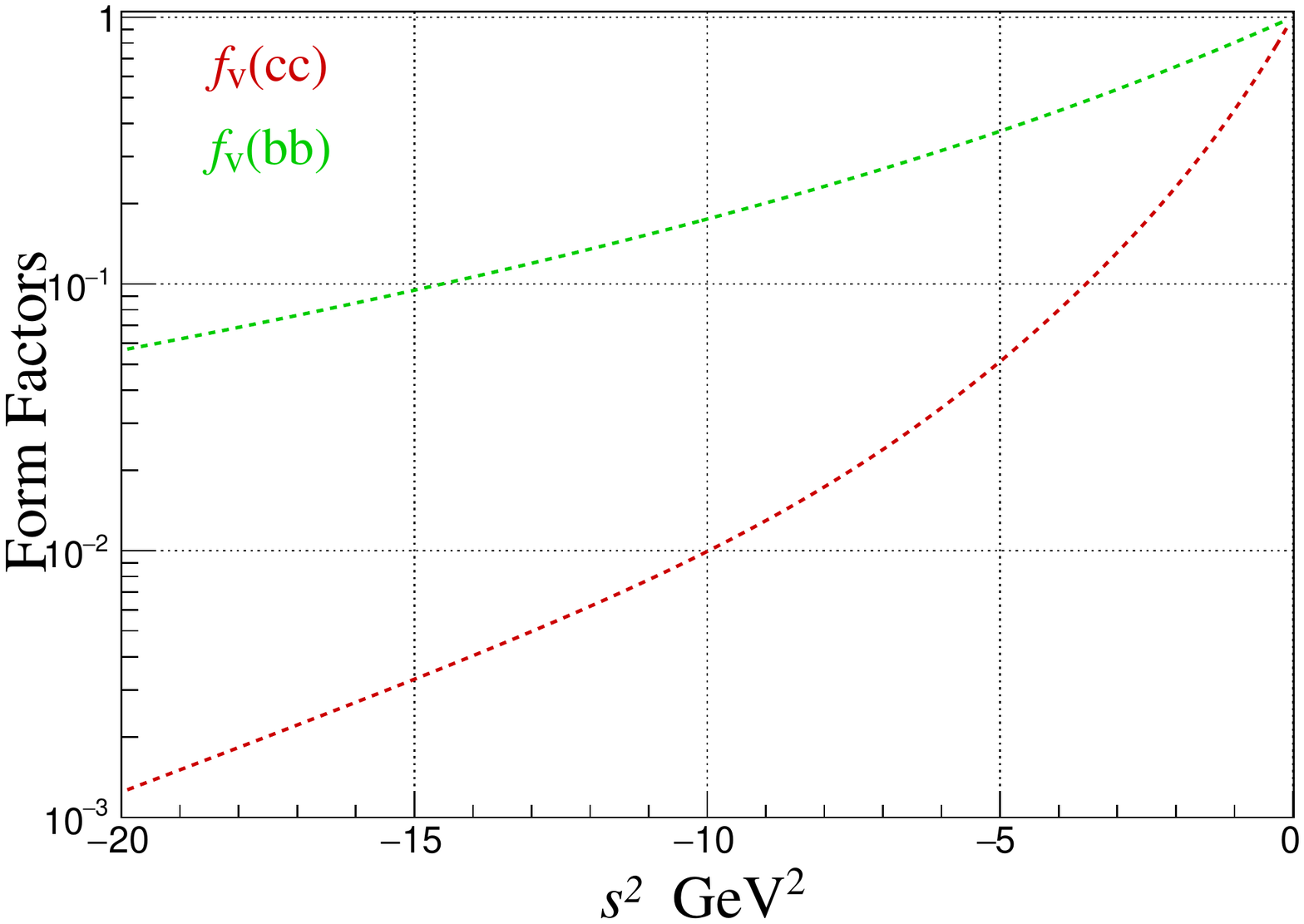} \label{Fig-form-II}}\\
\caption{Form factors of the $1^+$ diquarks $cc$ and $bb$. Method I denotes the results obtained by containing $V_\up{Conf}$ in the diquark interaction kernel while method II not.}\label{Fig-form-factors}
\vspace{0.5em}
\end{figure}

The diquark form factors $f_\up{V}(s^2)$ can be calculated by using the BS wave functions of the diquark. The attractive diquark is in the color antitriplet with the corresponding color factor $(-\frac{2}{3})$, which makes the bound interaction be half of the corresponding meson. Namely, in the rainbow-ladder truncation the effective interaction for the diquarks is reduced by a factor of 2 compared to that in the meson channel. Then after a charge conjugate transformation, a diquark fulfills the same Bethe-Salpeter equation with a meson system with only the interaction kernel halved and the parity flipped\,\cite{Cahill1987,Maris2002,Roberts2011}. Then by solving the BSE of the $J^P=1^-$ $Q\bar Q$ meson system, we can obtain the corresponding mass spectra and wave functions for the doubly heavy diquarks $cc$ and $bb$\,(see Ref.\cite{LiQ2020} for detailed calculations). The corresponding form factors $f_\up{V}$ describing the interaction between the diquark and a gluon have also been obtained in the previous work\,\cite{LiQ2020}, which are showed graphically in \autoref{Fig-form-I} labeled as method I. On the other hand, a colored diquark is never observed in nature alone and the confinement in diquark is an open problem in quark model. Therefore, we also calculated the corresponding form factors without considering the confinement item $V_\up{Conf}(s)$ in the Cornell potential of the diquark interaction kernel, and the obtained results are showed in \autoref{Fig-form-II} labeled as method II. Both form factors will be used in the calculation of the tetraquarks. The form factors fall faster with $s^2$ in method II than in method I. {\color{black} Also notice the form factors used here are calculated from the relevant BS wave functions combined with the Mandelstam formalism, which makes this work a self-contained framework and is different from the parameterized treatment  in Ref.\,\cite{KeHW2021}}.

{\color{black}Also notice the form factors in method I and II are calculated based on the on-shell diquark bound state, and then are simply generalized to the off-shell diquark propagator to describe the structure's non-pointlike effects. In order to investigate the influences of different form factors and make comparisons, we will also use another phenomenological form factors in this work, namely, 
\begin{gather} \label{E-form-propagator-type}
f_\up{V}(s^2) = \frac{\Lambda_c^2}{-s^2+\Lambda^2_c},
\end{gather}
where $\Lambda_c$ is a introduced regulator parameter. We notice that this propagator-like form factor is widely used in literature. Generally speaking, the parameter $\Lambda_c$ should be determined by fitting to the data. However, we found that this propagator-like function can not describe the obtained form factors well in a large range of the momentum transfer. Hence, in this work, we only fit in the small zone of $s^2$, namely, about $1(3)\sim0$\,GeV$^2$ for $cc\,(bb)$ diquark,  and the determined $\Lambda_c$ is $1.7$ and $2.8$\,GeV for the $cc$ and $bb$ diquarks respectively.  Also we will let $\Lambda_c$ change from $0.1$\,GeV to $20$\,GeV to investigate the corresponding dependence. 
}
\subsection{BSE under the instantaneous approximation}
Generally speaking, solving the four dimensional BSE \ref{E-BSE-T} is not an easy computational problem. However, under the instantaneous approximation introduced above, following Salpeter's procedures to cope with the fermion-antifermion system\,\cite{Salpeter1952}, we can reduce the four-dimensional BSE of the tetraquarks into the three-dimensional integral equation. 
Inserting the instantaneous kernel \ref{E-Kernel-3D} into the BSE \ref{E-BSE-T}, the tetraquark vertex can now be further expressed as
$
\Gamma^{\alpha\beta}(q) = \varrho_1 \varrho_2  \Theta^{\alpha\beta}(q_\perp) 
$, where the three-dimensional vertex $\Theta^{\alpha\beta}$ is defined as
\begin{align}
\Theta^{\alpha\beta}(q_\perp) &=\int \frac{\up d^3 k_\perp}{(2\pi)^3} \varkappa \left(k_\perp-q_\perp\right)  \varphi^{\alpha\beta}(k_\perp),
\end{align}
where the Salpeter wave function is defined by absorbing the integration over the time component as usual, 
\begin{gather}
\varphi^{\alpha\beta}(k_\perp) \equiv -i \int \frac{\up d k_P}{2\pi} T^{\alpha\beta} (k).
\end{gather}
The Salpeter wave function $\varphi^{\alpha\beta}(k_\perp)$ is only explicitly dependent on the three-dimensional momentum $k_\perp$.

Following the standard procedure, performing the contour integral over $q_P$ on both sides of \eref{E-BS-wave}, we obtain the three-dimensional Salpeter equation (SE),
\begin{align} \label{E-SE-1}
\varphi_{\alpha\beta}(q_\perp) &=d_{\alpha\mu}(p_{1\perp}) \Theta^{\mu\nu}(q_\perp) d_{\nu\beta}(p_{2\perp})\left(\frac{1 }{M-w_{1}-w_2}  - \frac{1}{M+w_{1}+w_2} \right).
\end{align}
Also we can define the positive and negative energy wave functions as 
\begin{gather}
\varphi^{\pm}_{\alpha\beta}(P,q_\perp) \equiv d_{\alpha\mu}(p_{1\perp}) \Theta^{\mu\nu}(q_\perp) d_{\nu\beta}(p_{2\perp})\frac{1 }{\pm M - (w_{1}+w_2)},
\end{gather} 
and we have $\varphi_{\alpha\beta}=\varphi_{\alpha\beta}^++\varphi_{\alpha\beta}^-$.  In the weak binding condition $M\sim (w_{1}+w_2)$, $\varphi_{\alpha\beta}^+\gg \varphi_{\alpha\beta}^-$, and the positive energy wave function $\varphi_{\alpha\beta}^+(q_\perp)$ dominates.  The SE can be further rewritten as the following simple Shr\"odinger type
\begin{align}
M^2 \varphi_{\alpha\beta} &=  (w_{1}+w_2)^2  \varphi_{\alpha\beta} +  2(w_1+w_2) d_{\alpha\mu}(p_{1\perp}) \Theta^{\mu\nu}(q_\perp) d_{\nu\beta}(p_{2\perp}). \label{E-SE-T}
\end{align}
The obtained three-dimensional BSE, namely, \eref{E-SE-T}, indicates that the mass of the tetraquark state consists of two parts, the kinetic energy and the potential energy. Also we notice \eref{E-SE-T} is in fact the integral equation of the Salpeter wave function $\varphi_{\alpha\beta}(q_\perp)$, and $M^2$ behaves as the eigenvalue of the corresponding Salpeter wave function. By solving this eigenvalue equation, we can obtain the mass spectra and wave functions of the corresponding tetraquarks.

The normalization condition of the Bethe-Salpeter wave function for the tetraquark state is generally expressed as,
\begin{align*}
i\int \frac{\up{d}^4 q}{(2\pi)^4}\frac{\up{d}^4 k}{(2\pi)^4}  \bar{T}_{\alpha\beta}(q,\xi')  \frac{\partial}{\partial P^0} \left[ \up{I}^{\alpha\alpha';\beta'\beta}(P,k,q) \right] T_{\alpha'\beta'}(k, \xi)  = (2M)\delta_{\xi \xi'},
\end{align*}
where the conjugate wave function is defined as $\bar T_{\alpha\beta} \equiv \gamma^0 T_{\alpha\beta}^\dagger \gamma_0$; the integral kernel reads,
\begin{align*}
\up{I}^{\alpha\alpha'; \beta'\beta}(P,k,q) = (2\pi)^4 \delta^4(k-q) [D_{\alpha\alpha'}(p_1)]^{-1}  [D_{\beta'\beta}(p_2)]^{-1} +i  K^{\alpha\alpha';\beta'\beta} (P,k,q).
\end{align*}
Both the inverses of the propagators and the interaction kernel $K$ are dependent on $P^0$ and $q_P$.
Inserting the inverses of the propagators, the integration involving the propagators behaves as,
\begin{equation}
\begin{aligned}
\up{N}_\up{propagator} &=\int \frac{\up d^3 q_\perp}{(2\pi)^3}\bar \varphi^{\alpha\beta} \Theta_{\alpha\beta} \frac{8Mw_1w_2}{M^2-(w_1+w_2)^2}.
\end{aligned}
\end{equation}
While the partial differential of the kernel gives 
\begin{gather}
\frac{\partial K^{\alpha\alpha';\beta'\beta}}{\partial P_0}=2(\alpha_1\varrho_2+\alpha_2\varrho_2) \varkappa g^{\alpha\alpha'}g^{\beta'\beta},
\end{gather} 
and then we obtain normalization related to the interaction kernel as,
 \begin{equation}
\begin{aligned}
\up{N}_\up{kernel} &=\int \frac{\up d^3 q_\perp}{(2\pi)^3}\bar \varphi^{\alpha\beta}  \Theta_{\alpha\beta}  \frac{4Mw_q}{(w_1+w_2)} ,
\end{aligned}
\end{equation}
where we have used the Salpter equation (\ref{E-SE-1}); $w_q \equiv \alpha_2 w_1+\alpha_1w_2$.
On the other hand, the vertex $\Theta_{\alpha\beta}$ can also be expressed by the wave function as,
\begin{gather}
\Theta_{\alpha\beta} =  \frac{M^2-(w_1+w_2)^2}{2(w_1+w_2)} \vartheta_{\alpha\mu}(p_{1\perp}) \varphi^{\mu\nu}  \vartheta_{\nu\beta}(p_{2\perp}).
\end{gather}
Putting the two part together and inserting the equation above, we obtain the normalization of the Salpeter wave function,
 \begin{equation} \label{E-Norm}
\int \frac{\up d^3 q_\perp}{(2\pi)^3} N_0\, \bar \varphi^{\alpha\beta}_{(\xi')}  \vartheta_{\alpha\mu} (p_{1\perp})  \vartheta_{\nu\beta}(p_{2\perp})   \varphi^{\mu\nu}_{(\xi)}  
=\delta_{\xi\xi'},
\end{equation}
where the abbreviation $N_0\equiv \left[\frac{2w_1w_2}{(w_1+w_2)}  + \frac{w_qM^2}{(w_1+w_2)^2}  -w_q \right]$ is used. By the dimensional analysis, we can conclude that the dimension of the Salpeter wave function $\varphi_{\alpha\beta}$ is $(-2)$ in units of mass.

{\color{black}
It should be noted that, the obtained three-dimensional (Bethe-)Salpeter equation \ref{E-SE-T} and the corresponding normalization condition \ref{E-Norm} are universal to any fully heavy tetraquark states consisting of two (axial)vector constituents, and do not depend on the specific properties of the total angular momentum $J$ or parity. These obtained equations are applicable to the Salpeter wave functions with $J^{PC}=0^{++}$, $1^{+-}$, $2^{++}$ or any other possible spin-parity. This is different from the approaches adopted in Ref.\,\cite{KeHW2021}, where the Salpeter equations are coupled with three certain wave functions.
The obtained equations and relevant results can also be further extended to deal with the molecular states consisting of two vector mesons.
}

\section{BS wave functions of the tetraquark states} \label{Sec-3}

A $1^+$ diquark and a $1^+$ antidiquark can form a boson with $J^P=0^+$, $1^+$, or $2^+$ in the ground state, namely, $\bm{1}\times \bm{1}=\bm{0}+\bm{1}+\bm{2}$. For the tetraquark state, the spatial parity is expressed as $P=(-1)^l$, with $l$ stands for the quantum number of the angular momentum.  A tetraquark state containing two $c$ quarks and two $\bar c$ quarks also occupies the definite $C$-parity. The charge conjugate parity\,($C$-parity) is expressed as $C=(-1)^{l+s}$ with $s$ stands for the spin of the particle. In order to simplify the expressions, from now on, we will always use the abbreviation $x_\alpha\equiv  \frac{q_{\perp\alpha}}{ |\vec q\,|}$. Since throughout this manuscript we work in the momentum space, this abbreviation is supposed not to cause confusion.

According to Lorentz condition, spin-parity and also considering the relativistic covariance, the Salpeter wave function of tetraquark states with $J^{P}=0^{+}$ can be generally expressed as,
	\begin{equation}
	\begin{aligned} \label{E-wave-0++}
	\varphi_{\alpha\beta}(q_\perp) &=  g_1\left( \hat P_\alpha \hat P_\beta -g_{\alpha\beta} \right)+g_2 x_{ \alpha} x_{\beta} ,
	\end{aligned}
	\end{equation}
where $\hat P_\alpha=\frac{P_\alpha}{M}$; the radial wave function $g_i(\vabs{q})~(i=1,\,2)$ just depends on $\vabs{q}$ explicitly. It is clear to see that $g_1$ corresponds to the $S$-wave component, and $g_2$ contributes to both the $S$ and $D$ partial waves (see Ref.\,\cite{LiQ2020} for a detailed expression in terms of the spherical harmonics $Y_l^m$).   
By inserting the two wave functions into \eref{E-Norm}, the normalization condition of the $0^{++}$ Salpeter wave function is obtained,
\begin{equation}\label{E-Norm-0}
\int \frac{\up{d}^3 q_\perp}{(2\pi)^3} N_0\left [n_0(g_1+g_2)^2+ 2g_1^2 \right]=1,
\end{equation}
where $n_0=\left(\frac{M_1M_2}{w_1w_2}\right)^2$.

Similarly, the $J^{PC}=1^{+-}$ Salpeter wave function can be constructed by using the the antisymmetric Levi-Civita tensor $\epsilon_{\alpha\beta \mu \nu}$ as
\begin{equation} 
\begin{aligned} \label{E-wave-1+-} 
\varphi_{\alpha\beta}(q_\perp) &= h_1 \epsilon_{\alpha\beta e \hat P}  + h_2 \epsilon_{\alpha\beta x \hat P } e\cdot x , 
\end{aligned}
\end{equation}
where $\epsilon_{\alpha\beta e\hat P }=\epsilon_{\alpha\beta \mu \nu} e^\mu \hat P^\nu $; $e_{\alpha}(\xi)$ is the polarization vector with $\xi\,(\xi=0,~\pm 1)$ denoting the possible polarization states, and fulfills the following conditions 
\begin{gather}
P_\alpha e^\alpha=0,\\
 \sum_\xi e_{\alpha}(\xi) e_\beta(\xi) = {G}_{\alpha\beta} \equiv\frac{P_\alpha P_\beta}{M^2} -g_{\alpha\beta}.
\end{gather}
It is clear to see that $h_{1}$ and $h_{2}$ parts in \eref{E-wave-1+-} represent the $S$ and $D$-wave components, respectively. {\color{red} }. The $J^{PC}=1^{+-}$ tetraquark wave function is antisymmetric under the interchange of the two free Lorentz index, which is different from the $0^{++}$ case. The $1^{+-}$ normalization is finally expressed as,
\begin{equation}\label{E-Norm-1}
\int \frac{\up{d}^3 q_\perp}{(2\pi)^3} \frac{2}{3}N_0\left [n_1h^2_1+(h_2-h_1)^2\right]=1,
\end{equation}
where $n_1={M_1^2}/{w_1^2} + {M_2^2}/{w_2^2}$. {\color{black} Notice in a similar work\,\cite{KeHW2021} only part of the $S$-wave components in the wave functions of $J^{PC}=0^{++}$ and $1^{+-}$ are included, while our results show that the $D$-wave components, namely, the $g_2$ and $h_2$ items,  and the the possible $S$-$D$ mixing effects also play important roles especially in the excited states\,(see the obtained wave functions in \autoref{Fig-wave-cccc}).}

The $J^{PC}=2^{++}$ Salpeter wave function of the tetraquarks can be constructed as
\begin{equation} 
\begin{aligned} \label{E-wave-2++}
\varphi_{\alpha\beta}(q_\perp) &=i_1 e_{\alpha\beta} + i_2 \left( \hat P_\alpha \hat P_\beta -g_{\alpha\beta} \right) e_{x x}+i_3 \left( e_{x\alpha} x_\beta + e_{x\beta}x_\alpha \right)+i_4 \left( x_{ \alpha} x_{ \beta} \right)e_{x x},
\end{aligned}
\end{equation}
where  $e_{x x}\equiv e_{\alpha\beta}x^\alpha x^\beta$; $e_{\alpha\beta}(\xi)$ is the symmetric polarization tensor with $\xi\,(=\pm2,\pm1,0)$ denoting the possible polarization states. The symmetric $e_{\alpha\beta}(\xi)$ is traceless, and also fulfills the Lorentz condition and the completeness relationship
\begin{gather}
e_{\alpha\beta}=e_{\beta\alpha},\\
e_{\alpha\beta}g^{\alpha\beta}=0,\\
P^\alpha e_{\alpha\beta}=0,\\
\sum_\xi e_{\alpha\beta}(\xi) e_{\alpha_1\beta_1}(\xi)=\frac{1}{2} \left(G_{\alpha \alpha_1}G_{\beta \beta_1}+G_{\alpha \beta_1}G_{\beta \alpha_1} \right)-\frac{1}{3}G_{\alpha \beta} G_{\alpha_1 \beta_1}. 
\end{gather}
It is clear to see that $i_{1}$,  $i_{2(3)}$, and $i_4$ parts represent the $S$ , $D$, and $G$-wave components, respectively. {\color{black}In Ref.\,\cite{KeHW2021} only the $i_1$ item representing the dominant $S$-wave component  is considered, while both the  $D$ and the possible $G$-wave components are ignored, which would have effects on the mass splittings and damage the completeness of the wave functions}.
The normalization for $2^{++}$ Salpeter wave function is expressed as,
\begin{align}\label{E-Norm-1}
\int \frac{\up{d}^3 q_\perp}{(2\pi)^3} &\frac{N_0}{15} \sum_{m=1,n\geqslant m}^4 c_{mn} i_m i_n =1
\end{align}
where the coefficients $c_{ij}$s are defined as
\begin{equation}
\begin{aligned}
          c_{11}&= 5n_0 -3c_0+10,  &c_{12}&= 4n_0-4,\\
          c_{13}&=- 14n_0 +6c_0-6, ~~&c_{14}&=4n_0,\\
          c_{22}&=2n_0+4,                &c_{23}&=-8n_0,\\
          c_{24}&=4n_0,                    &c_{33}&=11n_0 - 3c_0+3,\\
          c_{34}&=-8n_0,                   &c_{44}&=2n_0,
\end{aligned}
\end{equation}
with $c_0=(\frac{q^2}{w_1w_2})^2$.

Notice in the construction of the Salpeter wave functions, we work based on the good quantum number $J^{PC}$ while not the nonrelativistic characteristics, such as spin $S$ or orbital angular momentum $L$. Then the contributions from the $D$ or $G$-wave components are naturally included in the $0^{++}$, $1^{+-}$, and $2^{++}$ Salpeter wave functions, and are determined by the dynamics of the Bethe-Salpeter equation while not the man-made mixing effects. This is one of the advantages of the relativistic Bethe-Salpeter methods. The relativistic effects are naturally included in both the adoption of the BSE and the construction of the wave functions though the instantaneous approximation partly destroyed the covariance.  Inserting these constructed Salpeter wave functions into the three-dimensional Salpeter equation \ref{E-SE-T} and solving this eigenvalue problem numerically, we can obtain the corresponding mass spectra and wave functions, which are presented in following section.

\section{Numerical results and discussions} \label{Sec-4}

Before giving the numerical results of the mass spectra and wave functions, we specify the numerical values of the parameters first. The model parameters used in this work are kept the same with that we applied in previous meson and baryon calculations\,\cite{Chang2010,WangT2013,WangT2013A,LiQ2016,LiQ2017,LiQ2017A,LiQ2020},  
\[
a =e=2.7183,~~ \lambda =0.21~\si{GeV}^2,  ~~a_1 =a_2=0.06~\si{GeV};
\]
the strong interaction scale $ \Lambda_\text{QCD} =0.20\,\si{GeV}$ for $bb$ interaction while takes $0.27\,\si{GeV}$ for other cases; and the constituent quark masses used are $m_c =1.62~\si{GeV}, ~m_b=4.96~\si{GeV}$. The obtained constitute masses of the diquarks are $M_{cc}=3.303\,\si{GeV}$ and $M_{bb}=9.816\,\si{GeV}$\,\cite{LiQ2020} within method I, and $M_{cc}=3.135\,\si{GeV}$ and $M_{bb}=9.732\,\si{GeV}$ within the method II.

The free parameter $V_0$ plays a role in shifting the mass spectra and is determined by the spin-weighted average methods. By using the Clebsch–Gordan coefficients, the $J^P=0^+$  tetraquark formed by the  two $1^+$ diquark and antidiquarks can be decomposed as,
\begin{gather}
\textstyle \ket{(12)_1(34)_1}_0= \frac{1}{2} \ket{(14)_1(23)_1}_0 +\frac{\sqrt{3}}{2} \ket{(14)_0 (23)_0}_0,
\end{gather}
where label 1,\,2 denote the quarks, and 3,\,4 denote the antiquarks; $ \ket{(12)_1(34)_1}$ means quark-1 and quark-2 are in the spin-1 state, while the antiquark-3 and antiquark-4 are also in the spin-1 state; then other notations are also implied. Then the parameter $V_0$ for $0^+$ ${cc\bar c \bar c}$ will be expressed as $V_{0({cc\bar c\bar c})} = \frac14V_{0(J/\psi)} + \frac34V_{0(\eta_c)}$. By similar analysis,  the $J^P=1^+$  tetraquark is decomposed as,
\begin{gather}
\textstyle \ket{(12)_1(34)_1}_1= \frac{1}{\sqrt{2}}  \ket{(14)_0(23)_1}_1 + \frac{1}{\sqrt{2}}  \ket{(14)_1 (23)_0}_1 ,
\end{gather}
and then the corresponding parameter is determined as $V_{0({cc\bar c\bar c})} = \frac12\left[V_{0(J/\psi)} + V_{0(\eta_c)}\right]$. The $V_0$ for $2^+$ ${cc\bar c\bar c}$ is totally decided by the $V_{0(J/\psi)}$ since any two quarks (anitquarks) inside are in the spin-1 state. Finally, the obtained parameters $V_0$ for $0^{+}$, $1^{+}$, and $2^{+}$ $(cc\bar c\bar c)$ are $-0.304$, $-0.276$, and $-0.221$\,GeV respectively; for $bb\bar b\bar b$ are $-0.207$, $-0.192$, and $-0.162$\,GeV, respectively.
\begin{figure}[htbp]
\vspace{0.5em}
\centering
\subfigure[$(cc\bar c\bar c)$ in method I]{\includegraphics[width=0.315\textwidth]{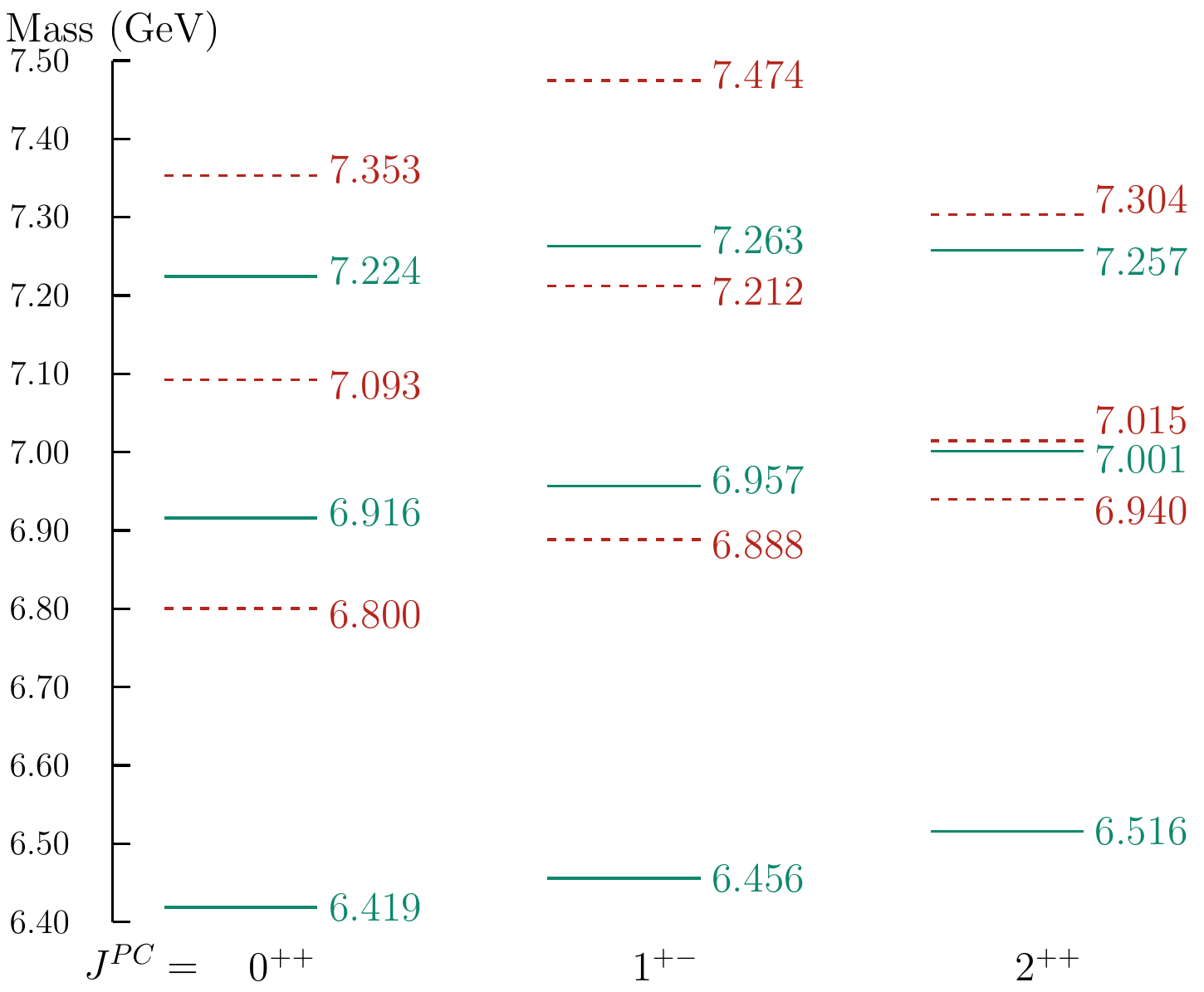} \label{Fig-cccc-spectra}}
\subfigure[$(cc\bar c\bar c)$ in method II]{\includegraphics[width=0.315\textwidth]{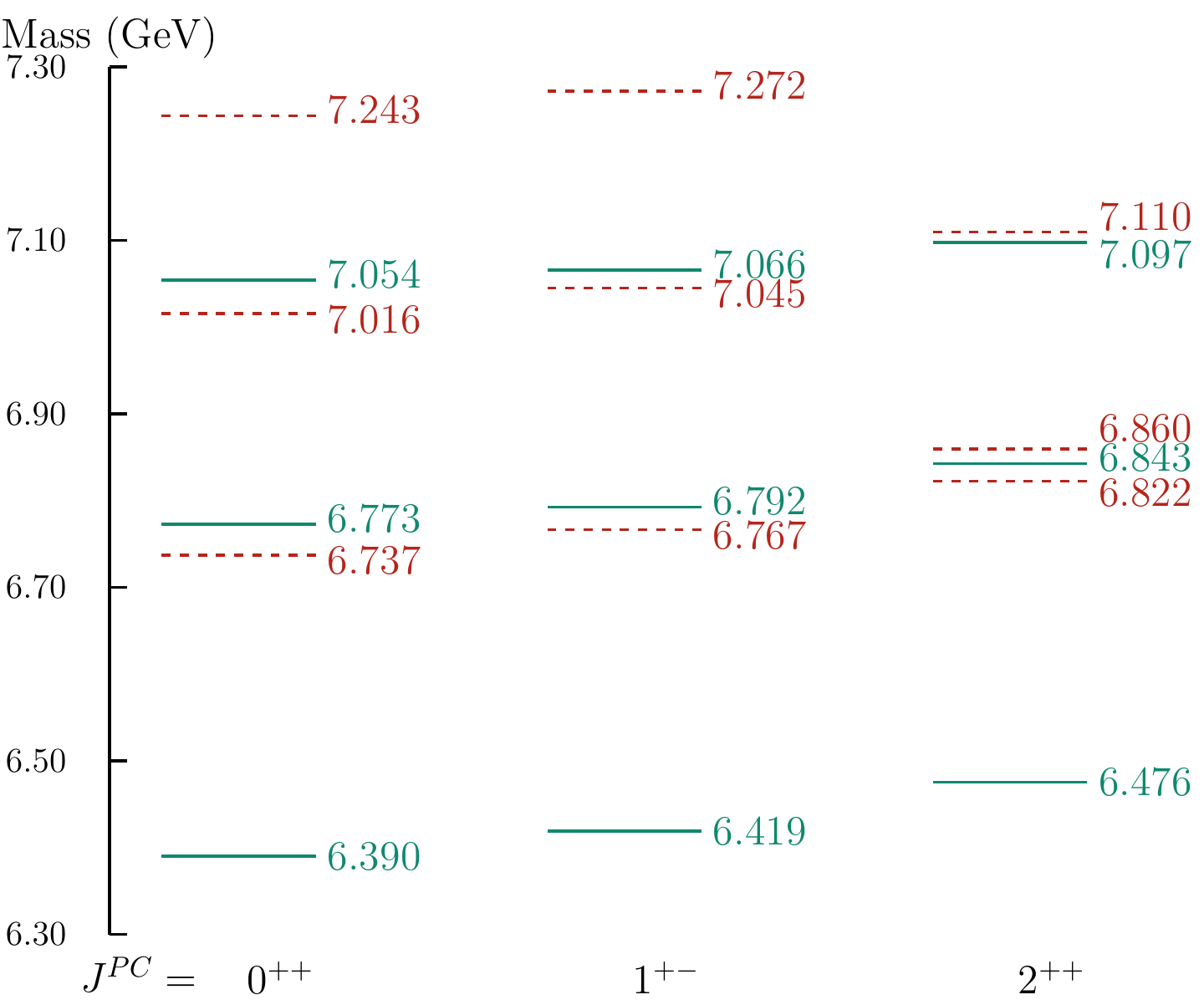} \label{Fig-cccc-spectra}} 
\subfigure[$(cc\bar c\bar c)$ in method III]{\includegraphics[width=0.315\textwidth]{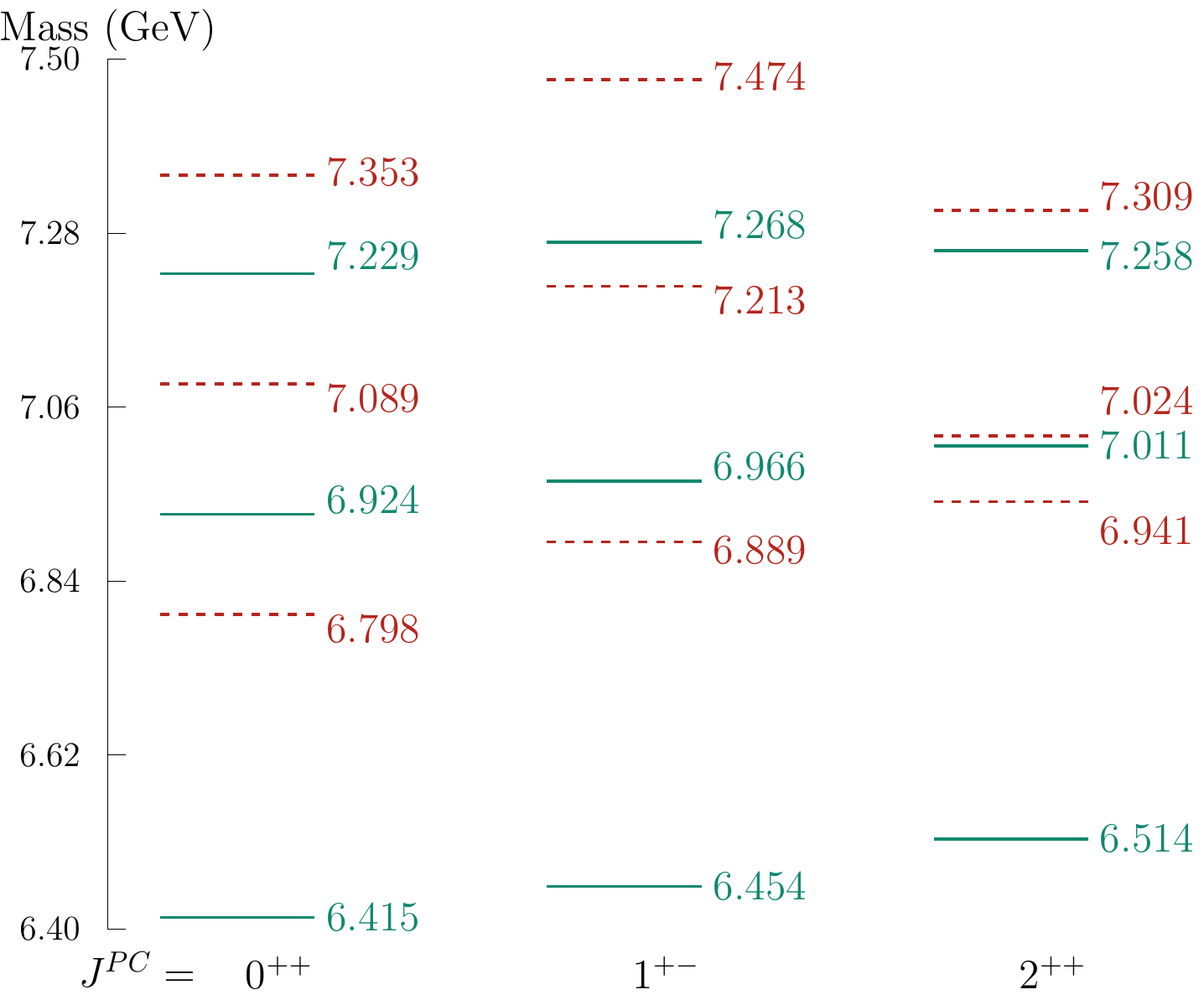} \label{Fig-cccc-spectra-Ac}} 
\subfigure[$(bb\bar b\bar b)$ in method I]{\includegraphics[width=0.315\textwidth]{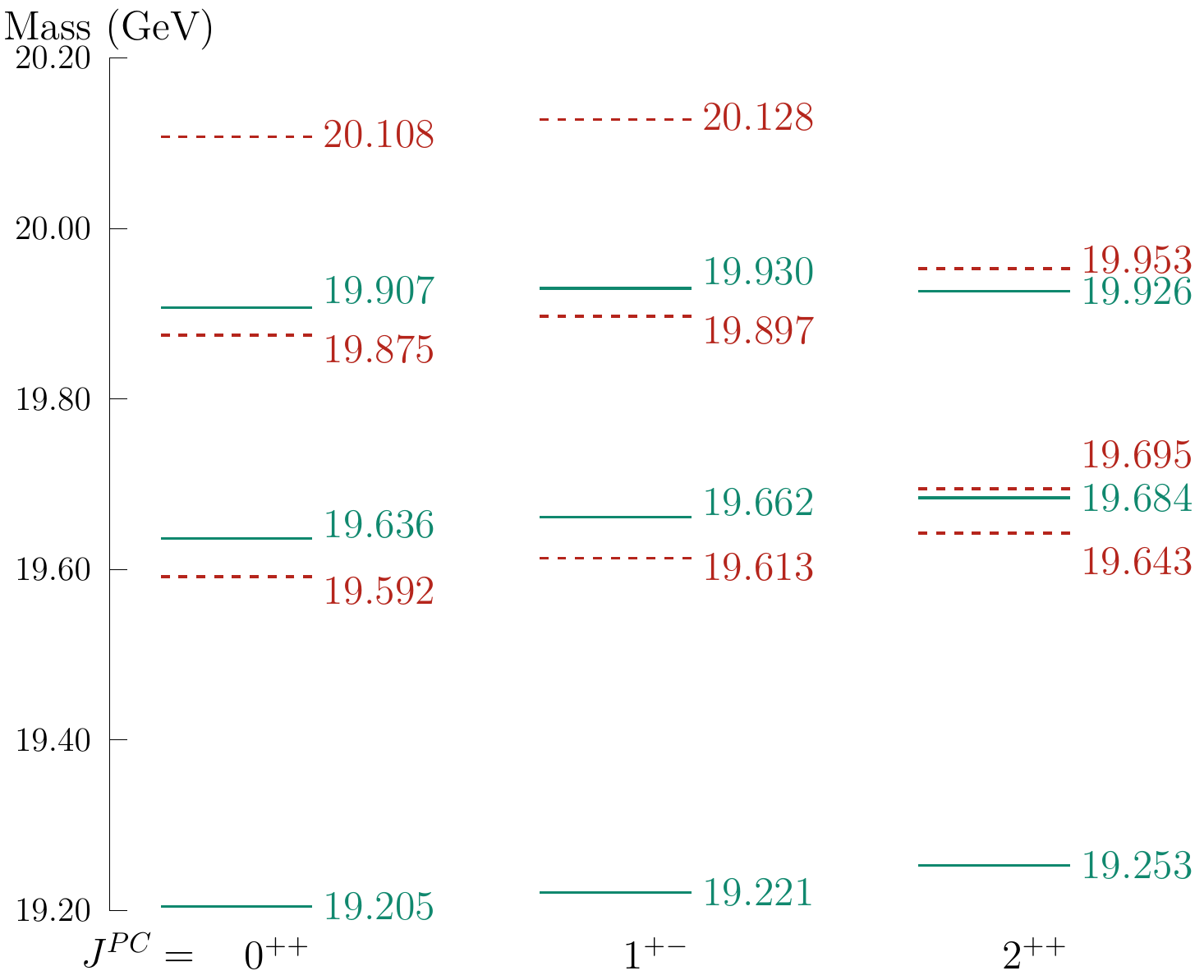} \label{Fig-bbbb-spectra}}
\subfigure[$(bb\bar b\bar b)$ in method II]{\includegraphics[width=0.315\textwidth]{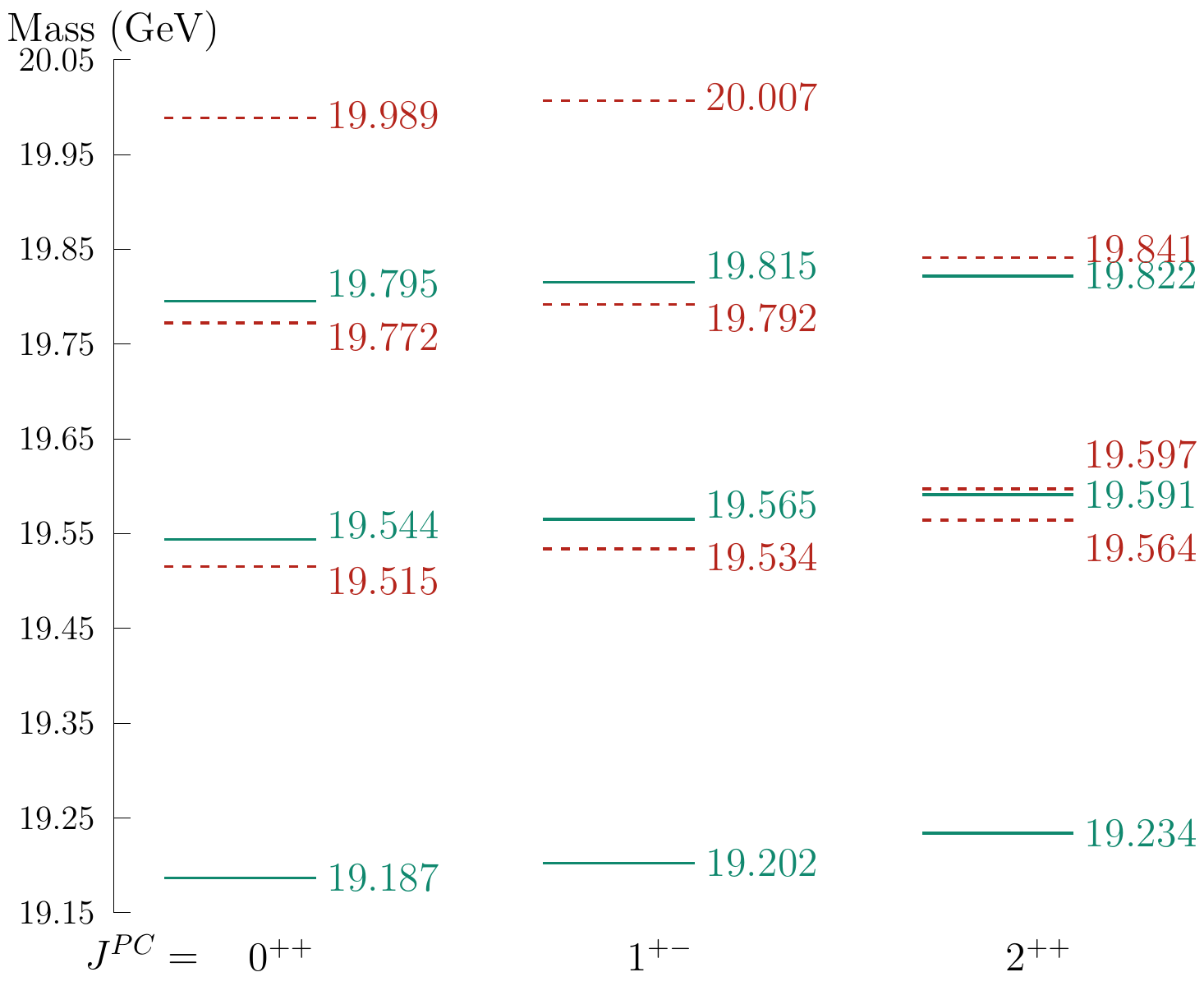} \label{Fig-bbbb-spectra}}
\subfigure[$(bb\bar b\bar b)$ in method III]{\includegraphics[width=0.315\textwidth]{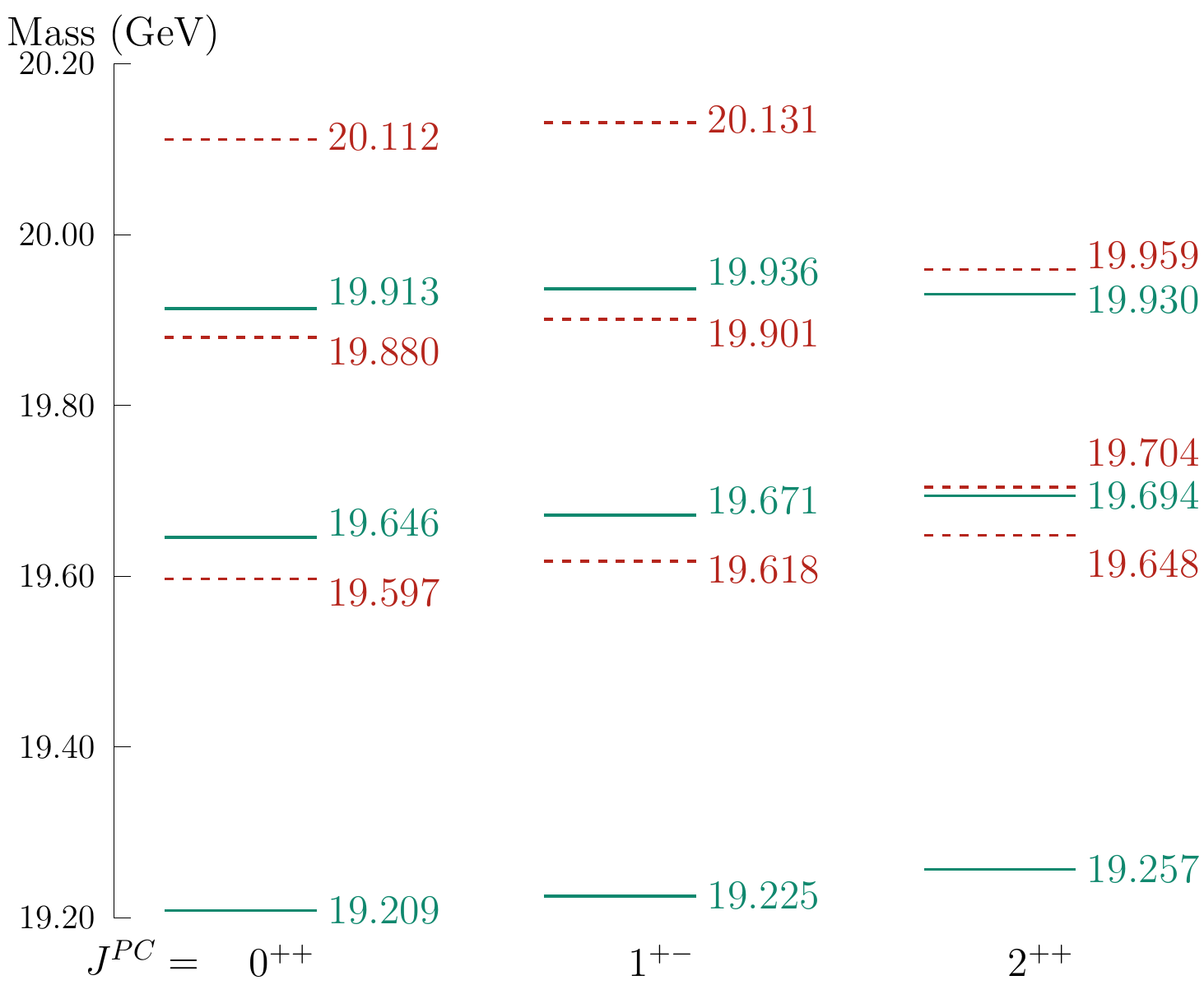} \label{Fig-bbbb-spectra-Ac}} 
\caption{Mass spectra of the tetraquark states $(cc\bar c\bar c)$ and $(bb\bar b\bar b)$ with all the diquarks\,(antiquarks) in the ground states. In method I\,(II) the diquark masses and corresponding form factors are calculated with (without) considering the confinement item $V_\up{Conf}$ in the interaction kernel; {\color{black} in method III the diquark form factors are assumed to be $\frac{\Lambda_c^2}{-s^2+\Lambda_c^2}$ with $\Lambda_c=1.7\,(2.8)$\,GeV for $cc\,(bb)$ diquark}.}\label{Fig-spectra}
\vspace{0.5em}
\end{figure}

\begin{figure}[h!]
\vspace{0.5em}
\centering
\subfigure[$(cc\bar c\bar c)$]{\includegraphics[width=0.48\textwidth]{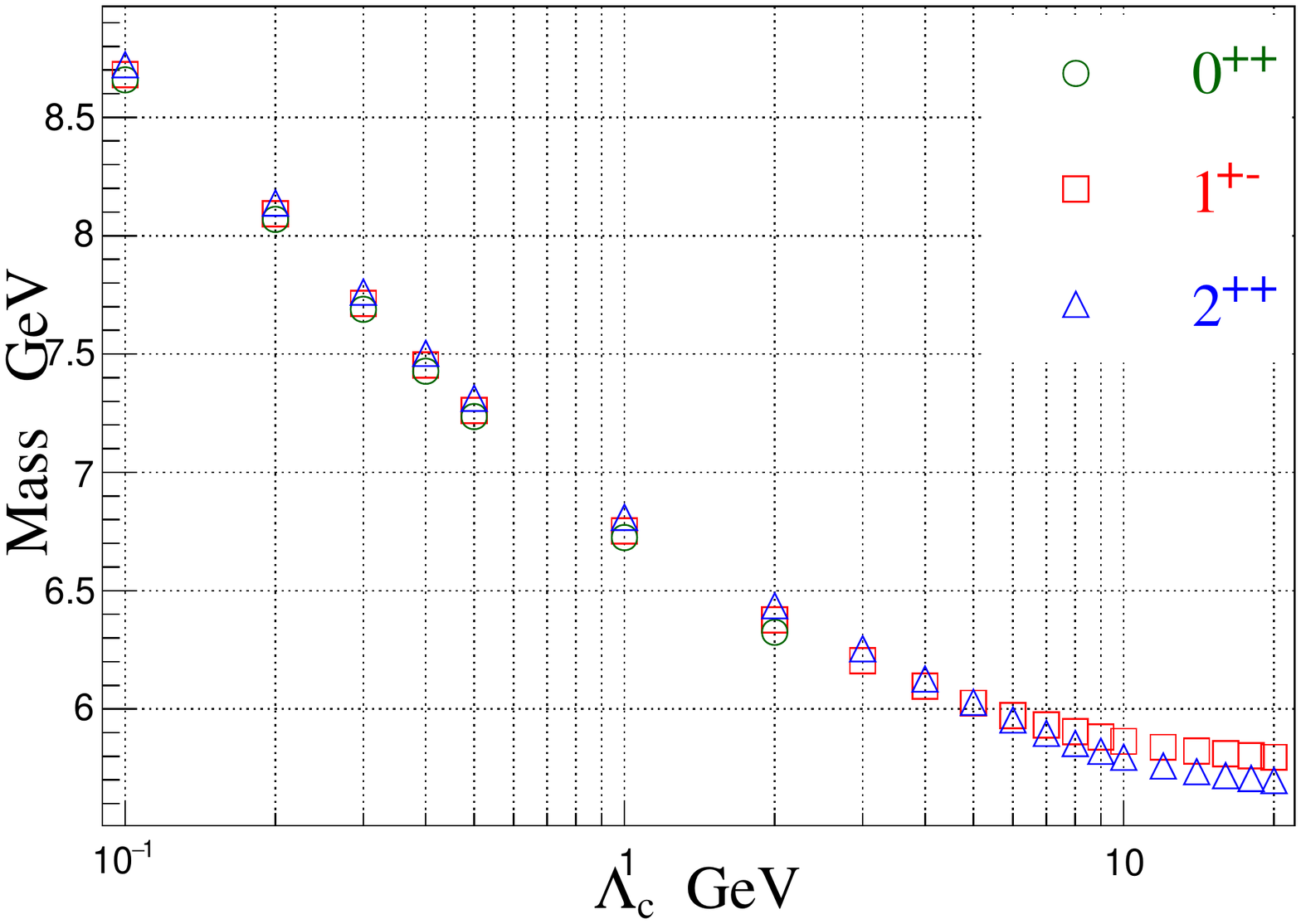} \label{Fig-cccc-spectra}}
\subfigure[$(bb\bar b\bar b)$]{\includegraphics[width=0.48\textwidth]{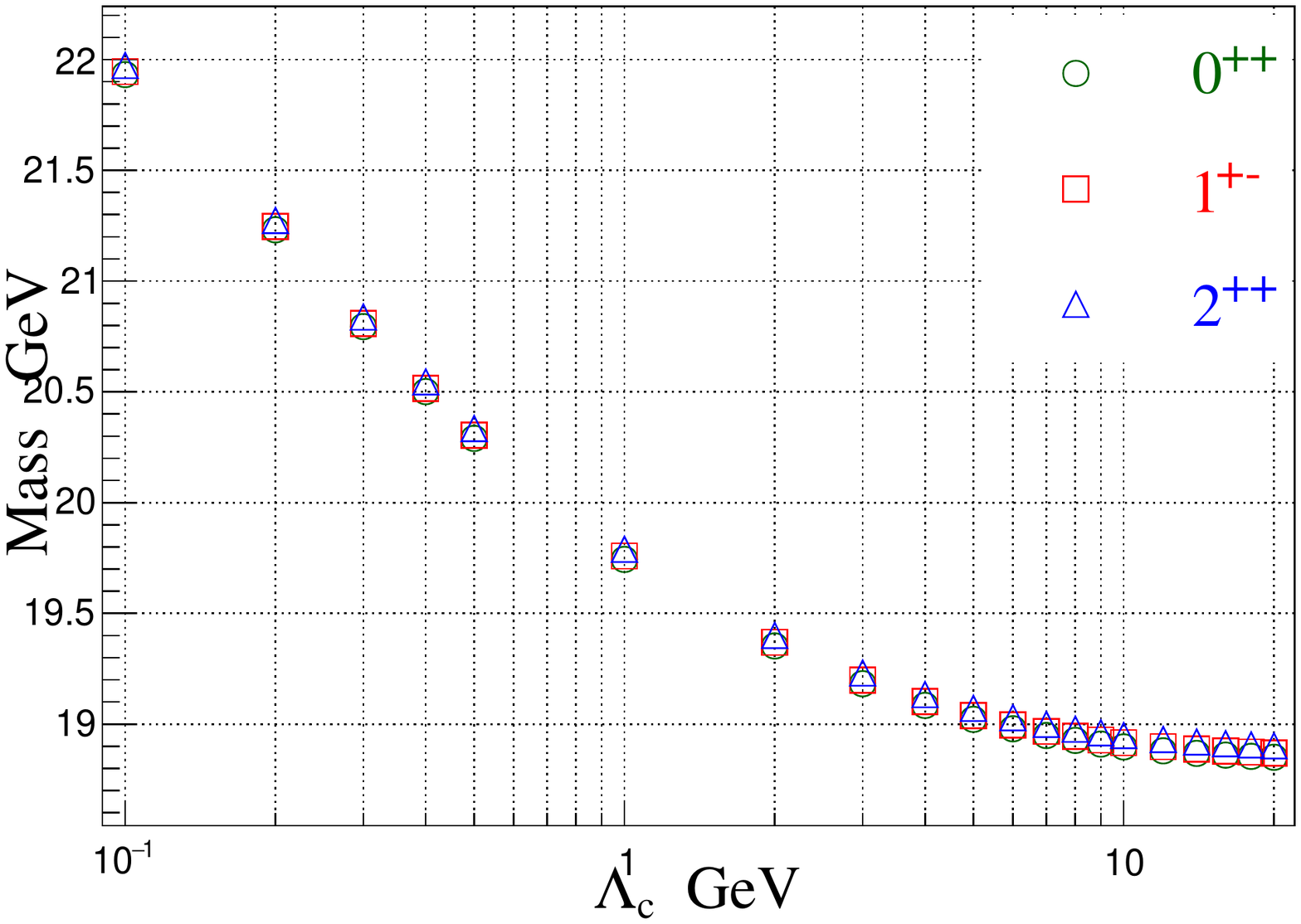} \label{Fig-cccc-spectra}} \\
\caption{ \color{black}Dependence on the regulator parameter $\Lambda_c$ of the tetraquark  mass spectra  in method III, where the diquark form factors are assumed to be $f_\up{V}(s^2)=\frac{\Lambda_c^2}{-s^2+\Lambda_c^2}$.}\label{Fig-Mass-Ac}
\vspace{0.5em}
\end{figure}

The obtained mass spectra of the tetraquarks $cc\bar c\bar c$ and $bb\bar b\bar b$ are showed in \autoref{Fig-spectra}, where method I denotes the results with considering the confinement item $V_\up{Conf}$ in calculating the diquark masses and the corresponding form factors, while method II not. {\color{black}The results labeled as method III represents the ones where the diquark form factors are assumed to be propagator-like type in \eref{E-form-propagator-type} with $\Lambda_c=1.7$\,GeV for $cc$ diquark and $2.8\,$GeV for $bb$ diquark respectively. In order to see the dependence on the regulator $\Lambda_c$ in method III, we show the variation of the mass spectra along with $\Lambda_c$ in \autoref{Fig-Mass-Ac} by changing the values of $\Lambda_c$ from 0.1\,GeV to 20\,GeV. With the increases of $\Lambda_c$, $f_\up{V}$ becomes more and more flat, which means the diquark is more and more similar to a pointlike particle.  Also it should be pointed out that the regulator parameter $\Lambda_c$ can not be taken too large which may cause the instability of the instantaneous BSE in high momentum zone. The plots reveal that the mass spectra in ground states are sensitive to the diquark form factors.} 
\begin{table}[htbp]
\caption{Comparison of the predicted $cc\bar c\bar c$ masses in ground states with $J^{PC}=0^{++}$, $1^{+-}$, and $2^{++}$  in units of GeV. }\label{Tab-Mass-cccc}
\vspace{0.2em}\centering
\begin{tabular}{ cccccccccccc }
\toprule[2pt]
$J^{PC}$ 		& This	  & \cite{DengCR2021} &\cite{LiuMS2019} &\cite{WangGJ2019} &\cite{LuQF2020}	&\cite{ChenW2017}		&\cite{ZhaoJX2020}	&\cite{Gordillo2020}	&\cite{Faustov2020}	&\cite{WengXZ2021}\\
\midrule[1.5pt]
$0^{++}$			&6.419 & 6.407,\,6.491			&6.470	&6.371	&6.435	&6.44		&6.346	&6.351	&6.190	&6.045\\
$1^{+-}$  			&6.456 & 6.463,\,6.580			&6.512	&6.450	&6.515	&6.37		&6.441	&6.441	&6.271	&6.231\\
$2^{++}$			&6.516 & 6.486,\,6.607			&6.534	&6.534	&6.543	&6.51		&6.475	&6.471	&6.367	&6.287\\
\bottomrule[2pt]
\end{tabular}
\end{table}
The obtained mass spectra show that the ground $cc\bar c\bar c$ tetraquarks locate in the range $6.4\sim6.5\,\si{GeV}$ when considering $V_\up{Conf}$ in diquarks. The mass splittings for the $1^{+-}$ and $2^{++}$ to $0^{++}$
states are about $35\,\si{MeV}$ and $100\,\si{MeV}$ respectively. When the confinement item is not included in the diquarks, the corresponding masses then locate about $40\,\si{MeV}$ lower. The first excited states are always about $400\,\si{MeV}$ higher than their corresponding ground states. A comparison of our predictions with recent researches is listed in \autoref{Tab-Mass-cccc}. Our obtained results for the three ground $cc\bar c\bar c$ states are roughly consistent with other studies. 
{\color{black} 
The mean values  in \autoref{Tab-Mass-cccc} are 6.36, 6.43, and 6.48\,GeV for the $0^{++}$, $1^{+-}$, and $2^{++}$ $cc\bar c\bar c$ in ground states respectively; and the corresponding standard derivations are 0.13, 0.10 and 0.09\,GeV respectively. All the results listed in \autoref{Tab-Mass-cccc} are above the threshold of the lowest quarkonium pair $\eta_c\eta_c$. Hence these three ground states are expected to be broad, since all of them can decay to a pair of quarkonia $\eta_c\eta_c$ or $J/\psi J/\psi$ through the (anti)quark rearrangements. These kind of decays are favored both dynamically and kinematically}.
From \autoref{Tab-Mass-cccc} we can conclude that the obtained masses of the ground $cc\bar c\bar c$ states are usually much lower (about $400\sim500$ MeV) than the $X(6900)$ observed by the LHCb collaboration. The observed $X(6900)$ is less likely to be the ground state of the compact tetraquark $cc\bar c \bar c$ states, but might be the first or second radial excited states. However, more detailed information is needed to investigate the inner structure of $X(6900)$.  {\color{black} We also notice that the masses of the $cc\bar c\bar c$ in ground states is near to the $X(6900)$ with $\Lambda_c\sim 1$\,GeV in method III. }

The obtained $bb\bar b\bar b$ masses in ground states are in the range $19.2\sim 19.3$ GeV, which are higher than the $\Upsilon\Upsilon$ threshold $1.892$\,GeV and $\eta_b\eta_b$ threshold $1.880$\,GeV but lower than the $\chi_{b0}\chi_{b0}$ threshold\,\cite{PDG2014}. Also we notice that $M_{bb}$ in method II is about 80\,MeV lower than that in method I, while the obtained mass spectra of $bb\bar b\bar b$ are about 20\,MeV lower in ground states. The comparison of our predictions with other researches is collected in \autoref{Tab-Mass-bbbb}, from which we can see that the theoretical masses of $bb\bar b\bar b$ in ground states locate in a large range of $18.75\sim 19.35\,\si{GeV}$ in researches.
 {\color{black} 
The mean values of the ground $bb\bar b\bar b$ tetraquarks in \autoref{Tab-Mass-bbbb} are about 19.14, 19.20, and 19.22\,GeV for the $0^{++}$, $1^{+-}$, and $2^{++}$, respectively, where the corresponding standard derivations are 0.20, 0.18 and 0.16\,GeV respectively. Except the 3 results predicted in Ref.\,\cite{Bedolla2020} and the $0^{++}$ state in Ref.\,\cite{WengXZ2021}, all other results listed in \autoref{Tab-Mass-bbbb} are higher than the $\Upsilon\Upsilon$ threshold. Therefore, the three states can decay to a pair of quarkonia $\eta_b\eta_b$ or $\Upsilon\Upsilon$ through the quark rearrangements, and hence are expected to be broad. 
}
\begin{table}[htbp]
\caption{Comparison of the $bb\bar b\bar b$ masses in ground states with $J^{PC}=0^{++}$, $1^{+-}$, and $2^{++}$  in units of GeV.  }\label{Tab-Mass-bbbb}
\vspace{0.2em}\centering
\begin{tabular}{ cccccccccccc }
\toprule[2pt]
$J^{PC}$ 		& This	   &\cite{LuQF2020}	 &\cite{ZhaoJX2020}	&\cite{Faustov2020}	&\cite{LiuMS2019}	&\cite{WengXZ2021}	&\cite{Gordillo2020}	&\cite{WangGJ2019}	&\cite{Bedolla2020}	\\
\midrule[1.5pt]
$0^{++}$	&19.205 & 19.201	& 19.154	& 19.314	& 19.322	& 18.836	& 19.199	&19.243	& 18.748\\
$1^{+-}$  	&19.221 & 19.251	& 19.214	& 19.320	& 19.329	& 18.969	& 19.276	&19.329	& 18.828\\
$2^{++}$	&19.253 & 19.262	& 19.232	& 19.330	& 19.341	& 19.000	& 19.289	&19.325	& 18.900\\
\bottomrule[2pt]
\end{tabular}
\end{table}

\begin{figure}[h!]
\vspace{0.5em}
\centering
\subfigure[$0^{++}(n=1)$]{\includegraphics[width=0.315\textwidth]{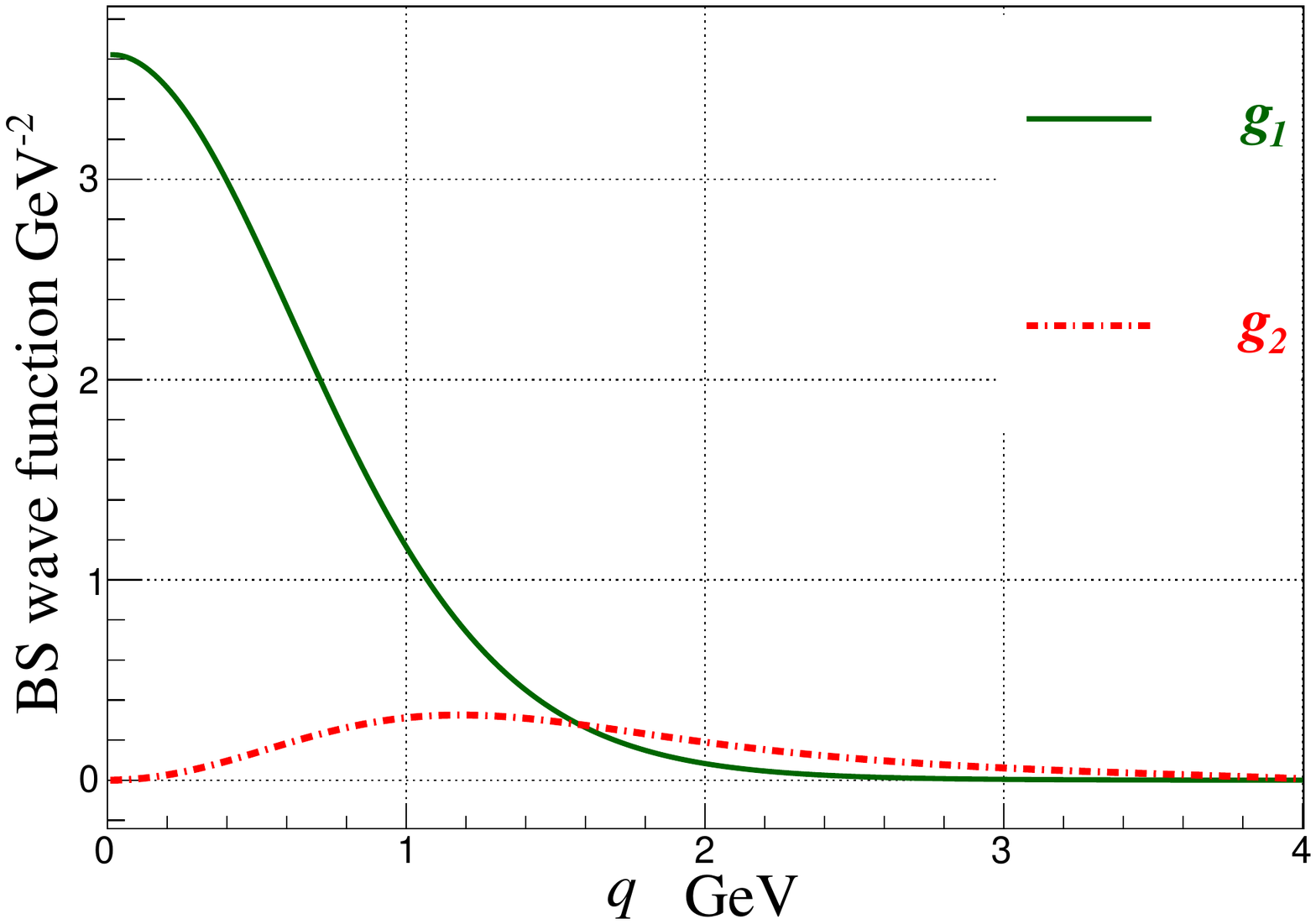} \label{Fig-wave-0-n1}}
\subfigure[$1^{+-}(n=1)$]{\includegraphics[width=0.315\textwidth]{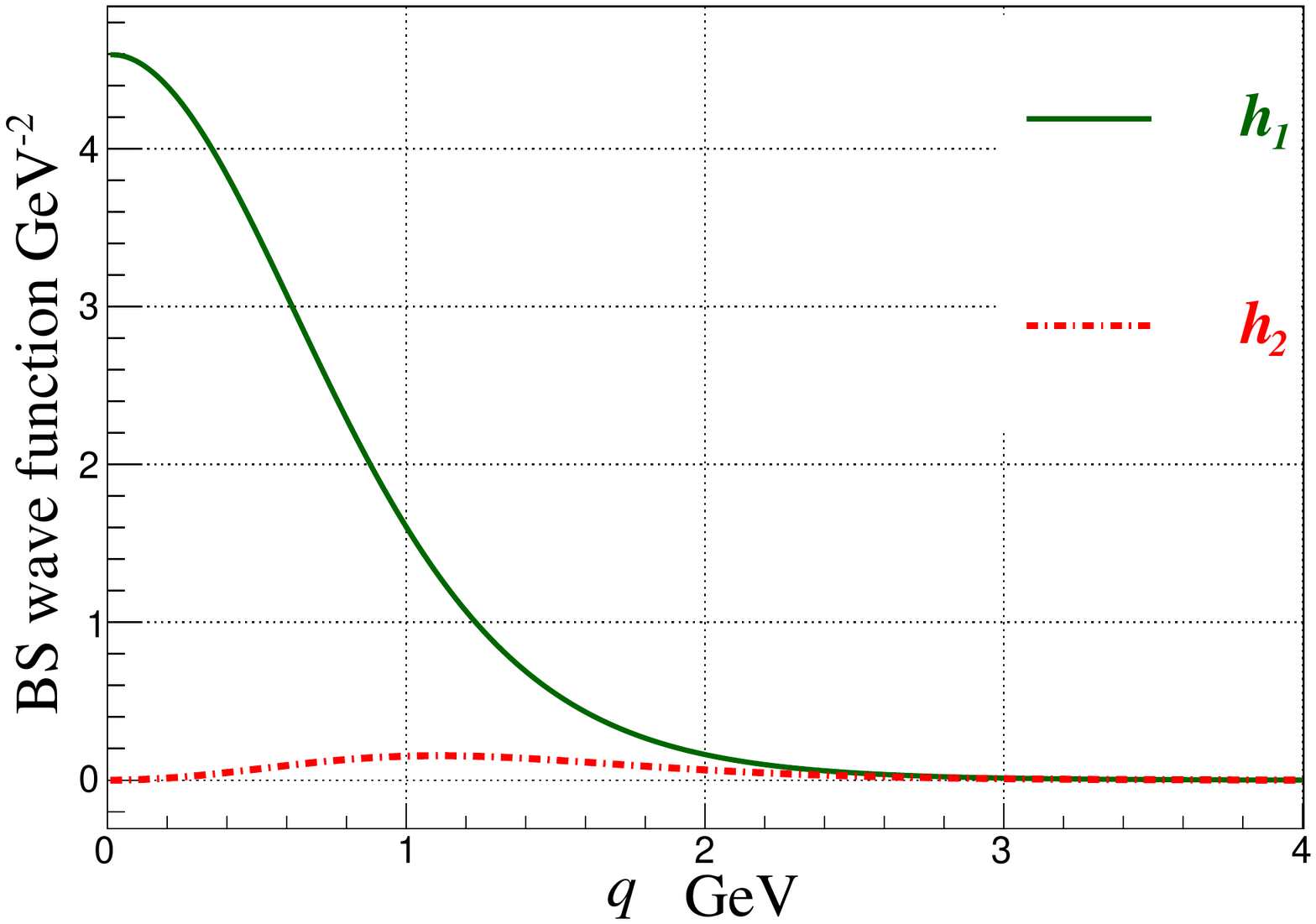} \label{Fig-wave-1-n1}}
\subfigure[$2^{++}(n=1)$]{\includegraphics[width=0.315\textwidth]{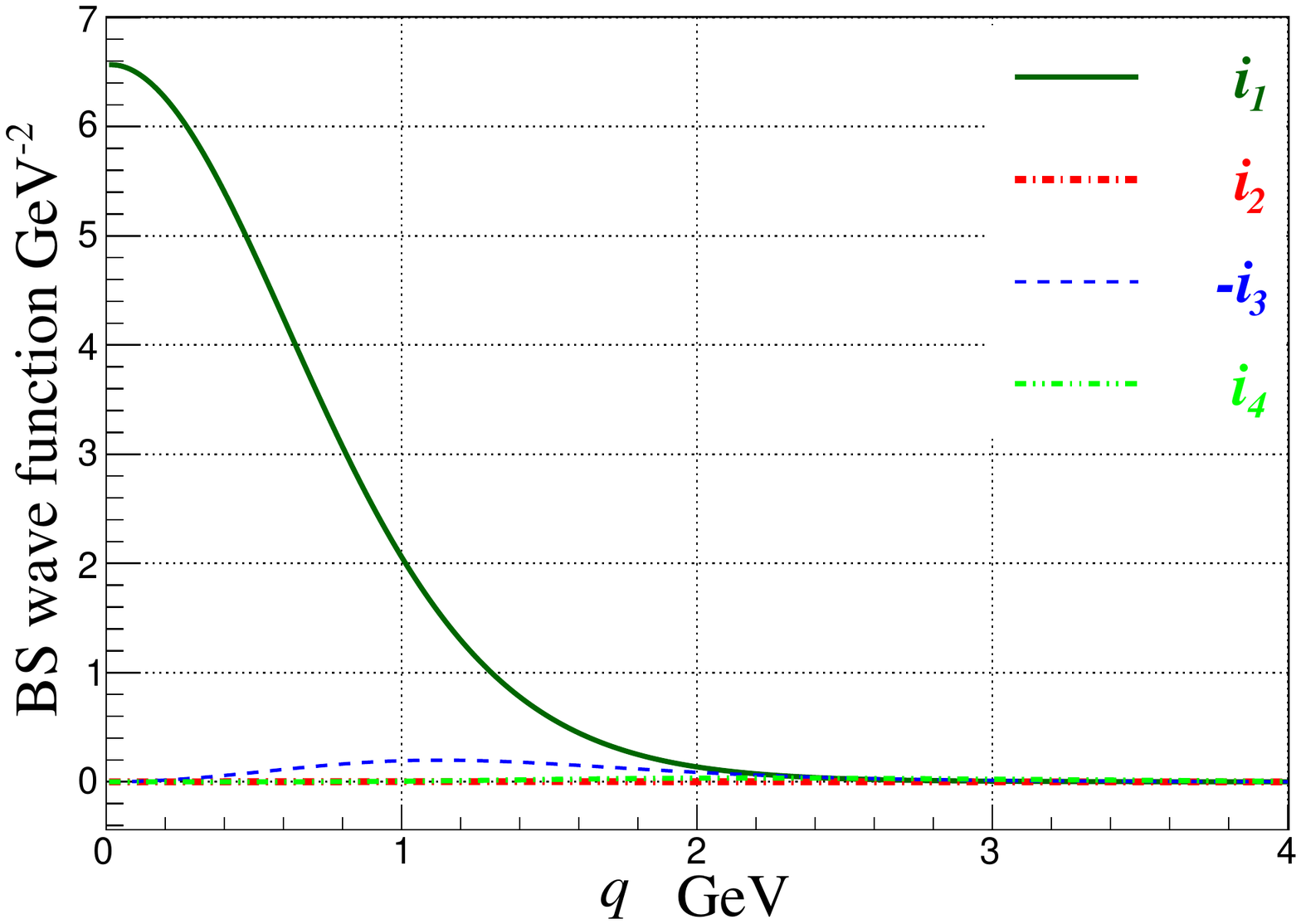} \label{Fig-wave-2-n1}}\\
\subfigure[$0^{++}(n=2)$]{\includegraphics[width=0.315\textwidth]{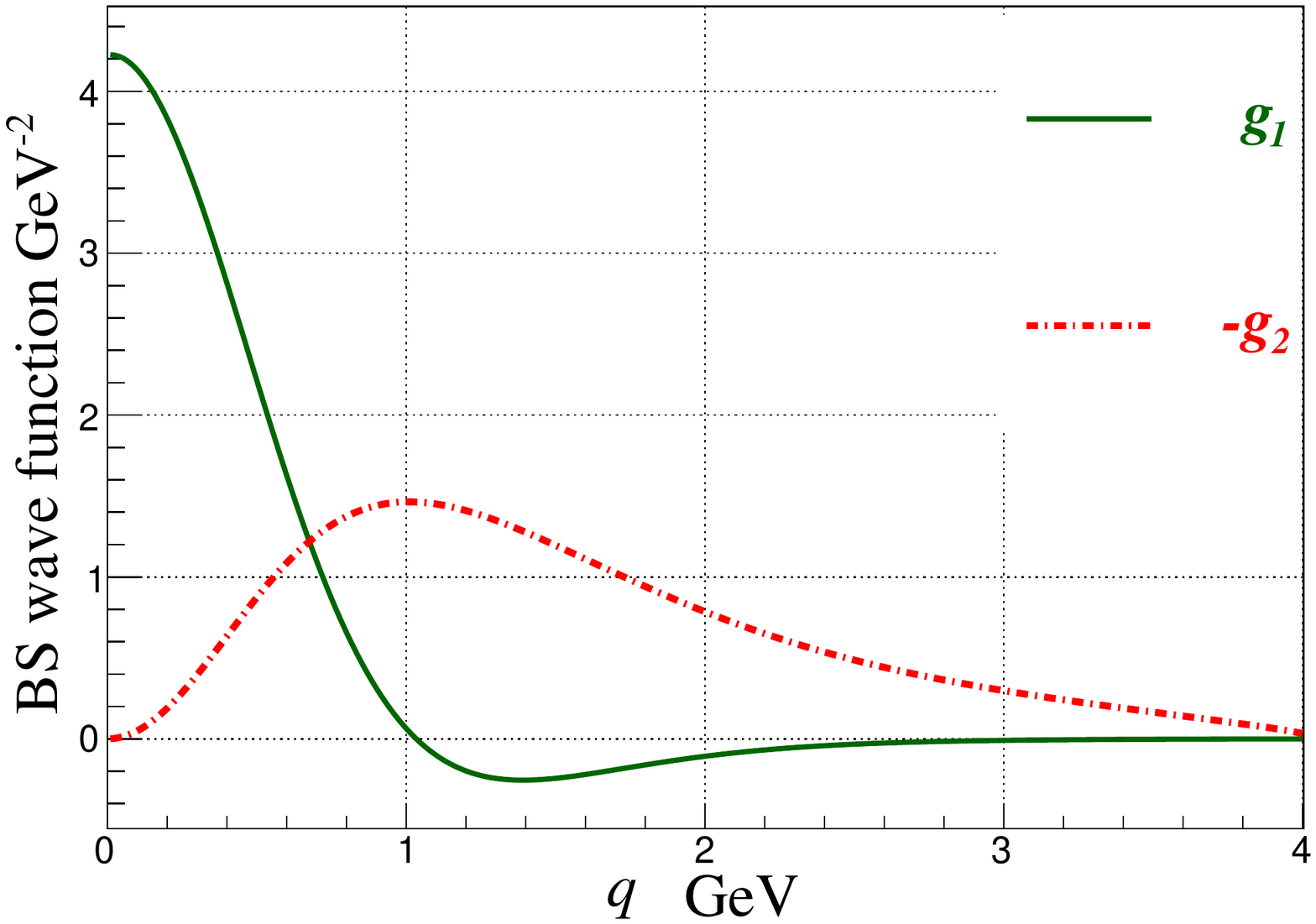} \label{Fig-wave-0-n2}}
\subfigure[$1^{+-}(n=2)$]{\includegraphics[width=0.315\textwidth]{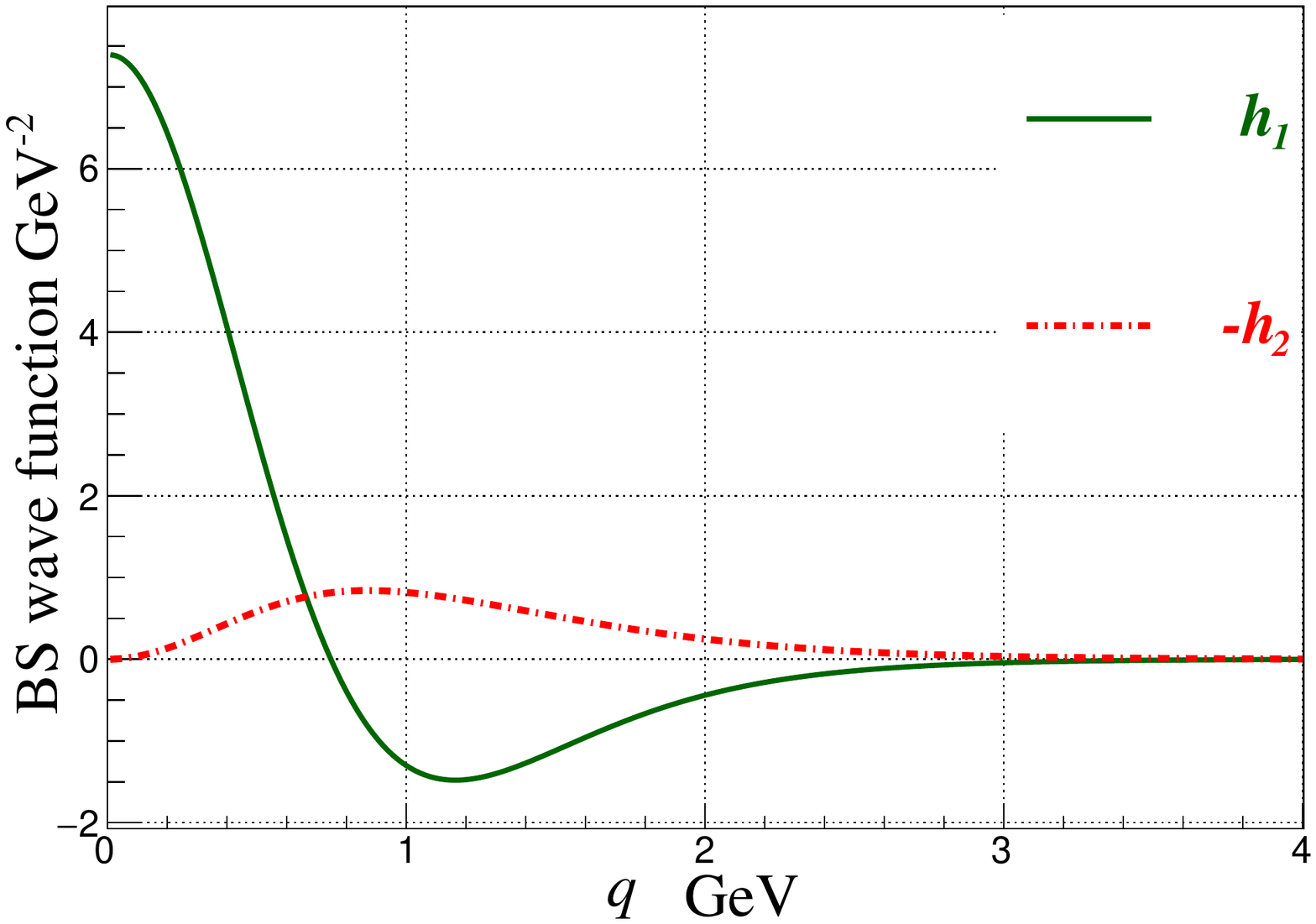} \label{Fig-wave-1-n2}}
\subfigure[$2^{++}(n=2)$]{\includegraphics[width=0.315\textwidth]{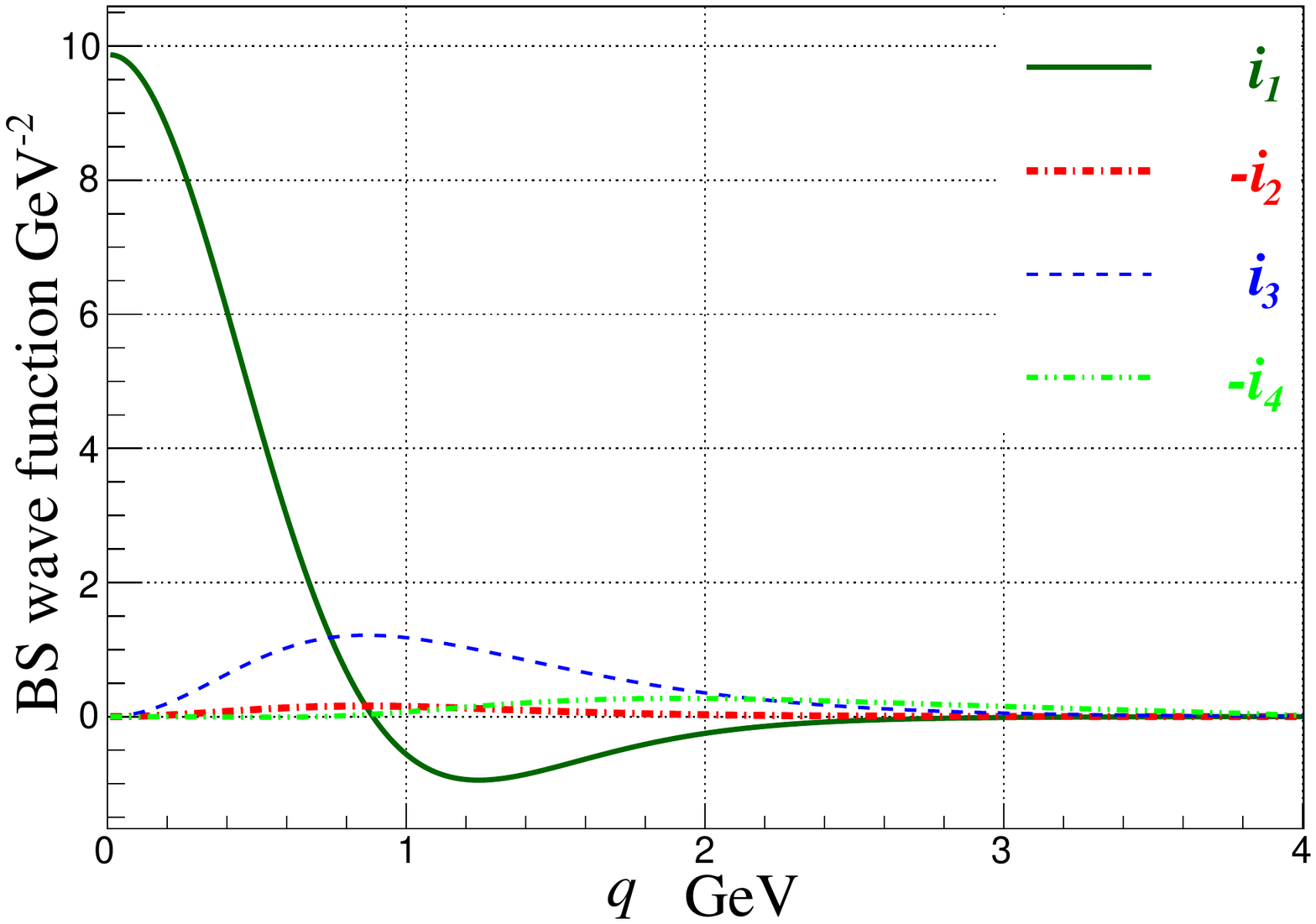} \label{Fig-wave-2-n2}}\\
\subfigure[$0^{++}(n=3)$]{\includegraphics[width=0.315\textwidth]{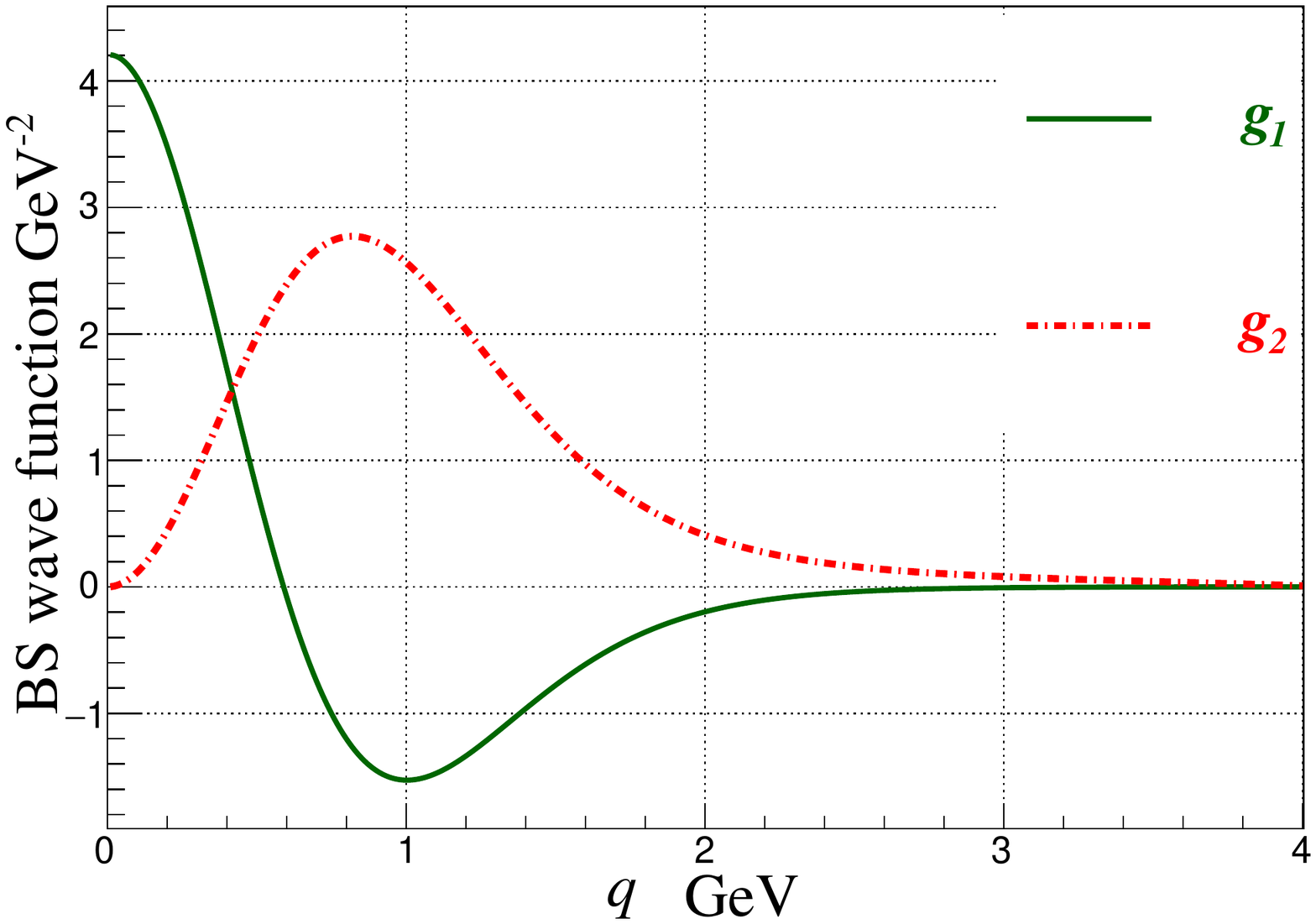} \label{Fig-wave-0-n3}}
\subfigure[$1^{+-}(n=3)$]{\includegraphics[width=0.315\textwidth]{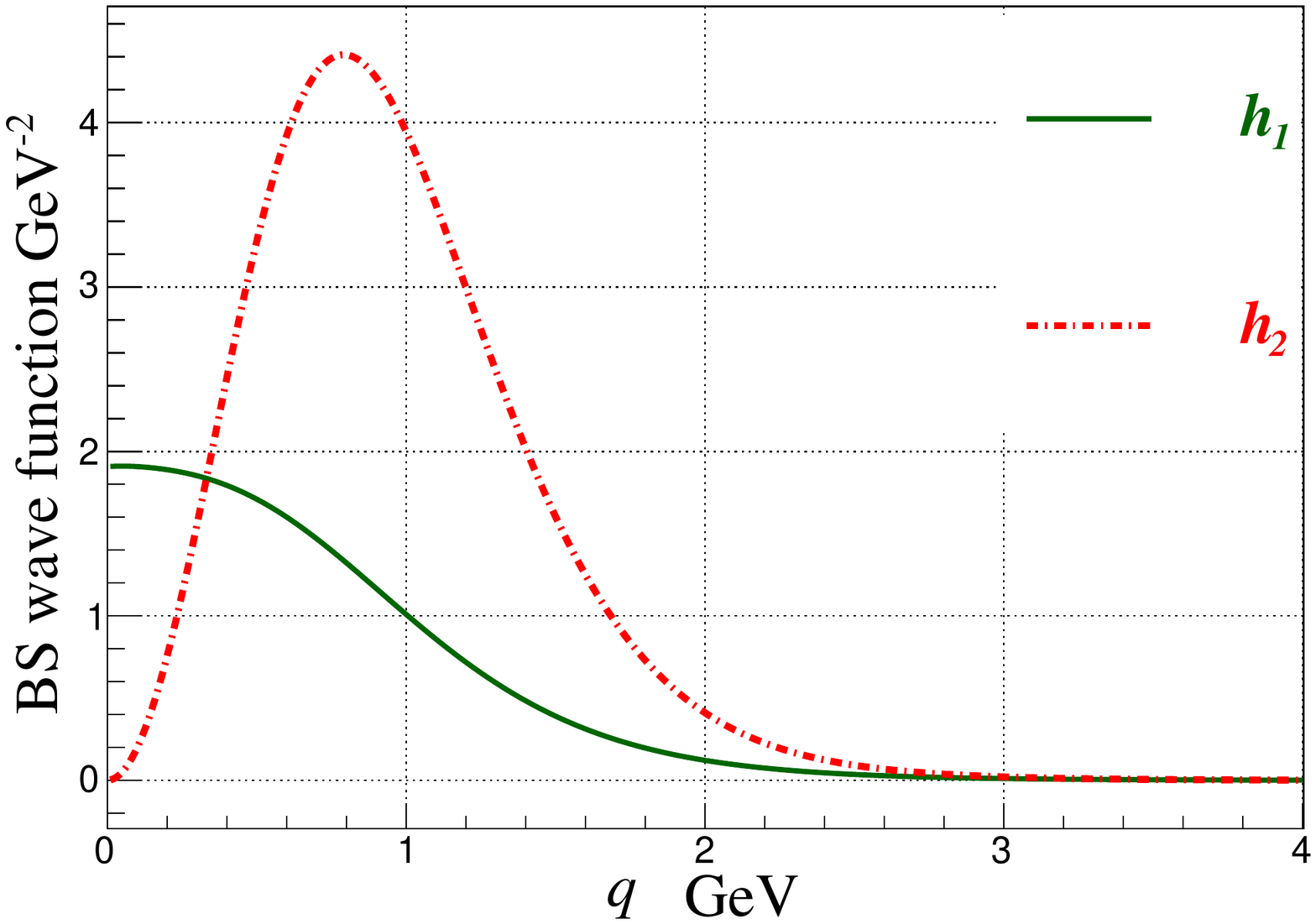} \label{Fig-wave-1-n3}}
\subfigure[$2^{++}(n=3)$]{\includegraphics[width=0.315\textwidth]{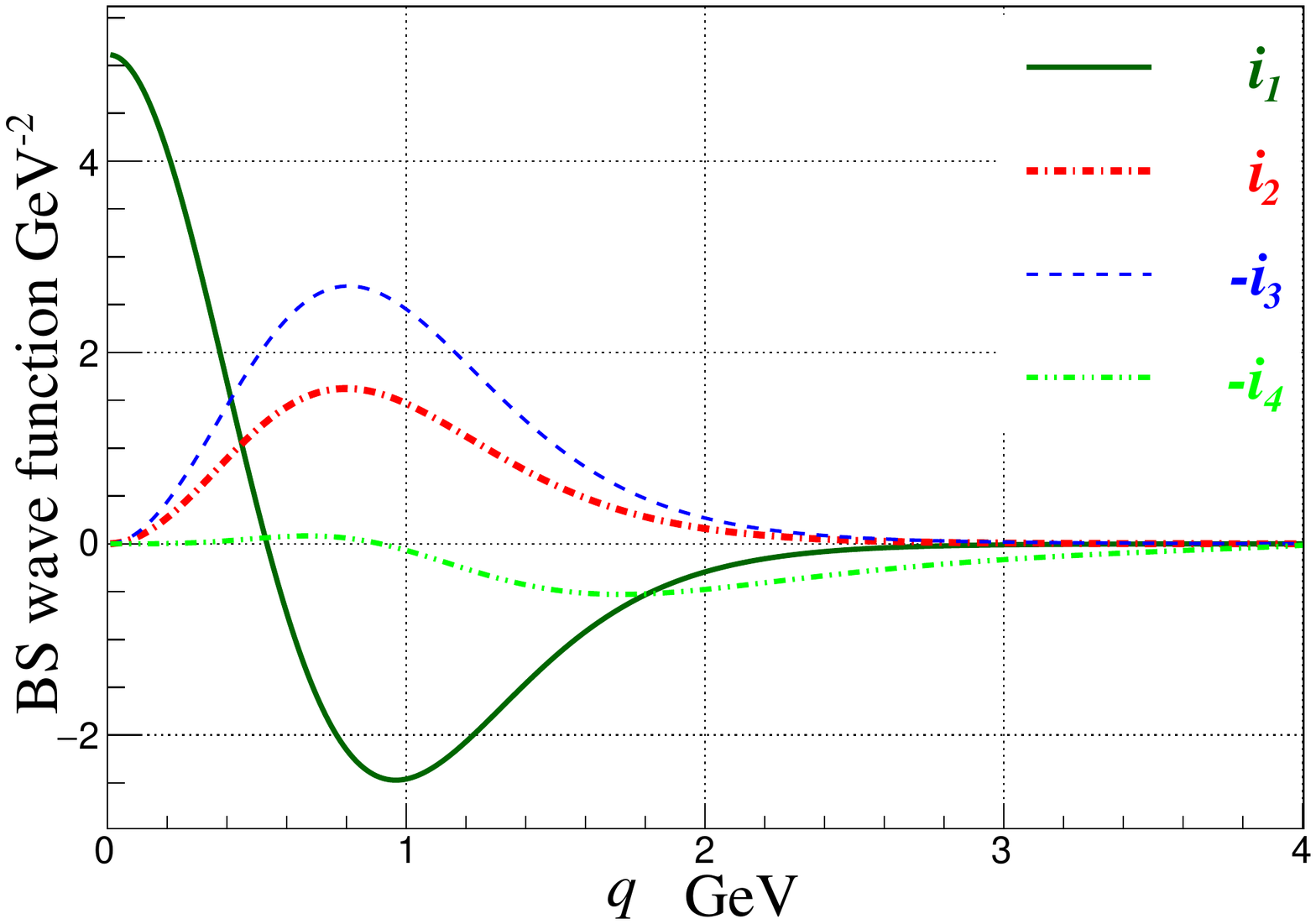} \label{Fig-wave-2-n3}}\\
\subfigure[$0^{++}(n=4)$]{\includegraphics[width=0.315\textwidth]{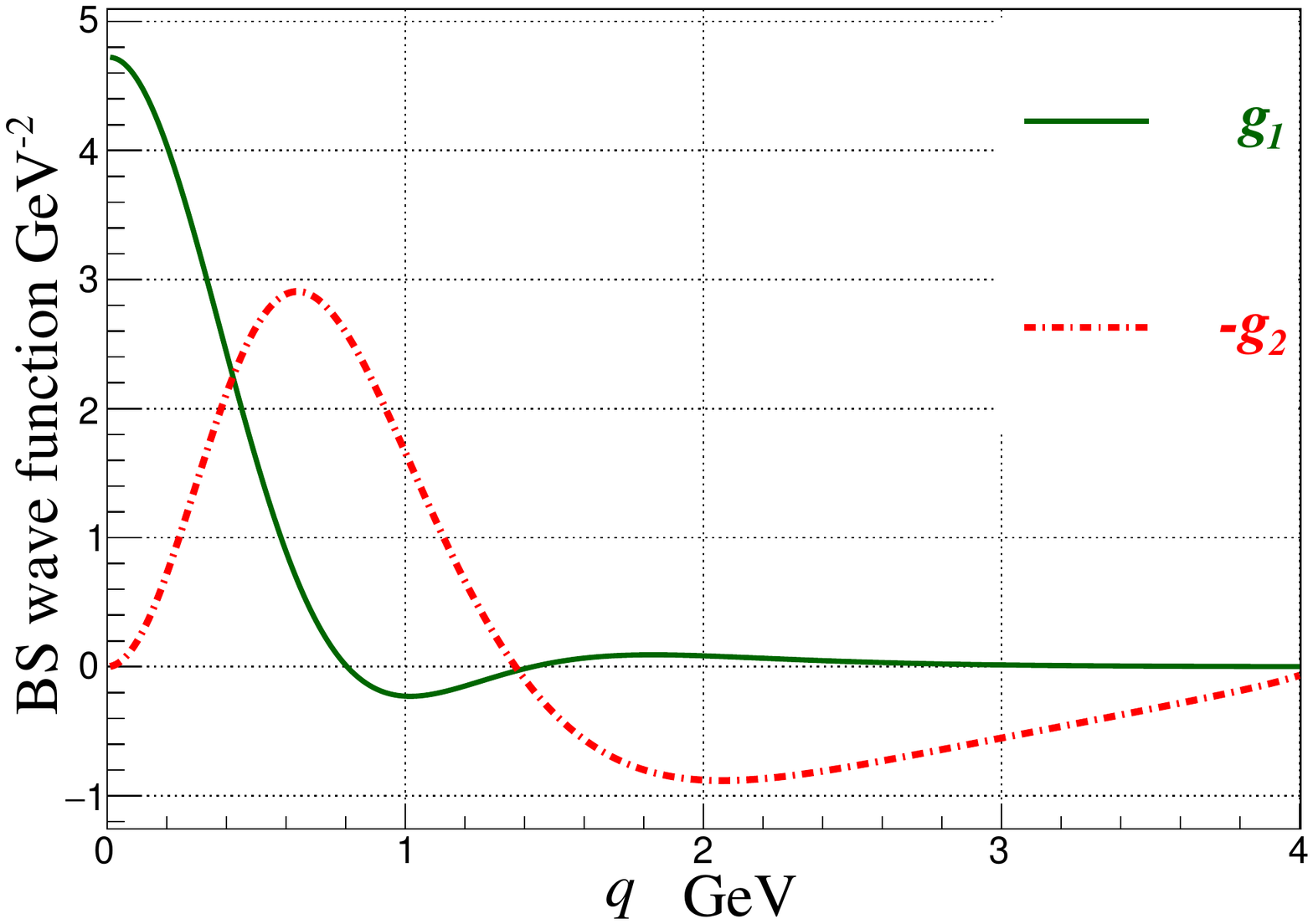} \label{Fig-wave-0-n4}}
\subfigure[$1^{+-}(n=4)$]{\includegraphics[width=0.315\textwidth]{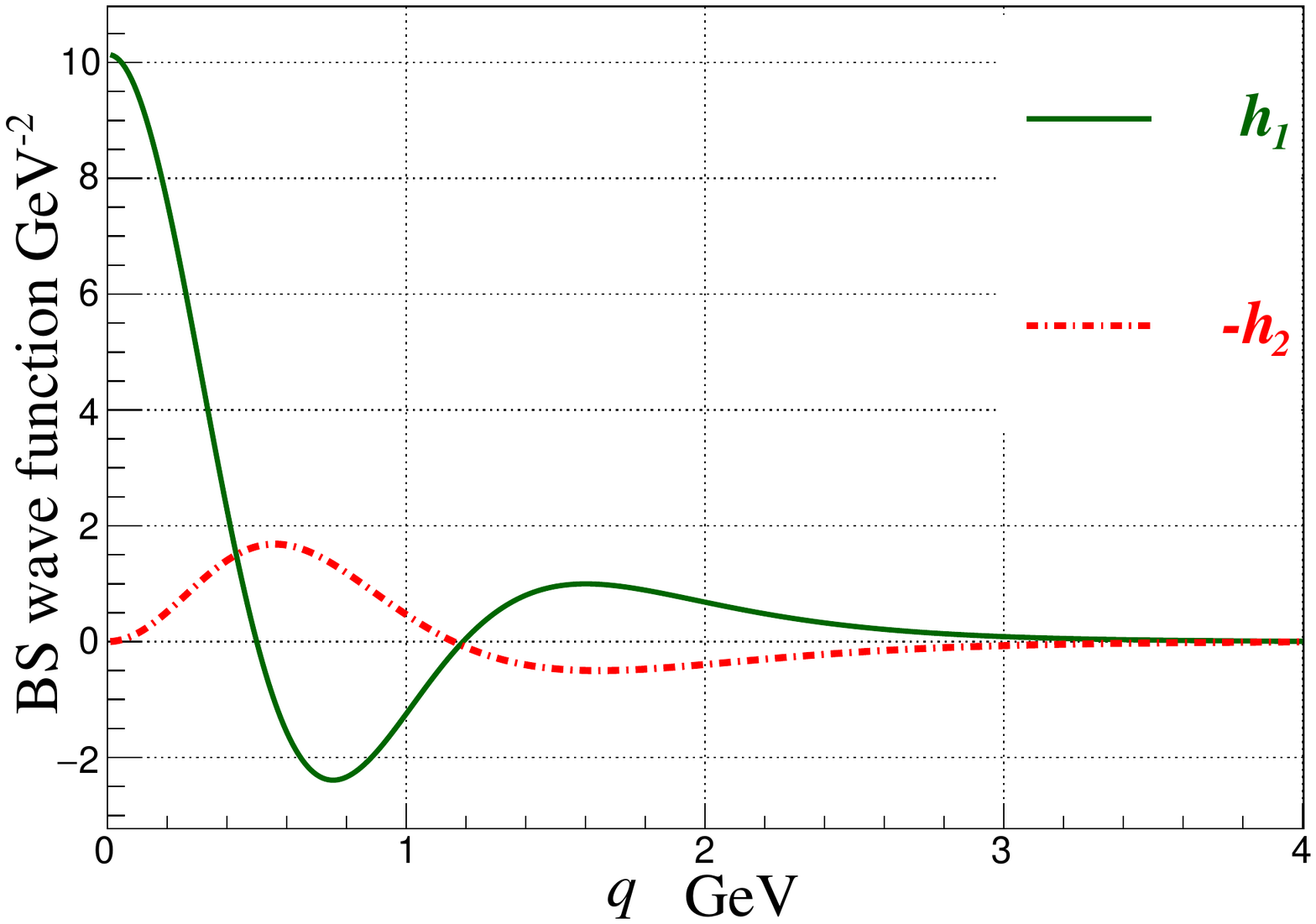} \label{Fig-wave-1-n4}}
\subfigure[$2^{++}(n=4)$]{\includegraphics[width=0.315\textwidth]{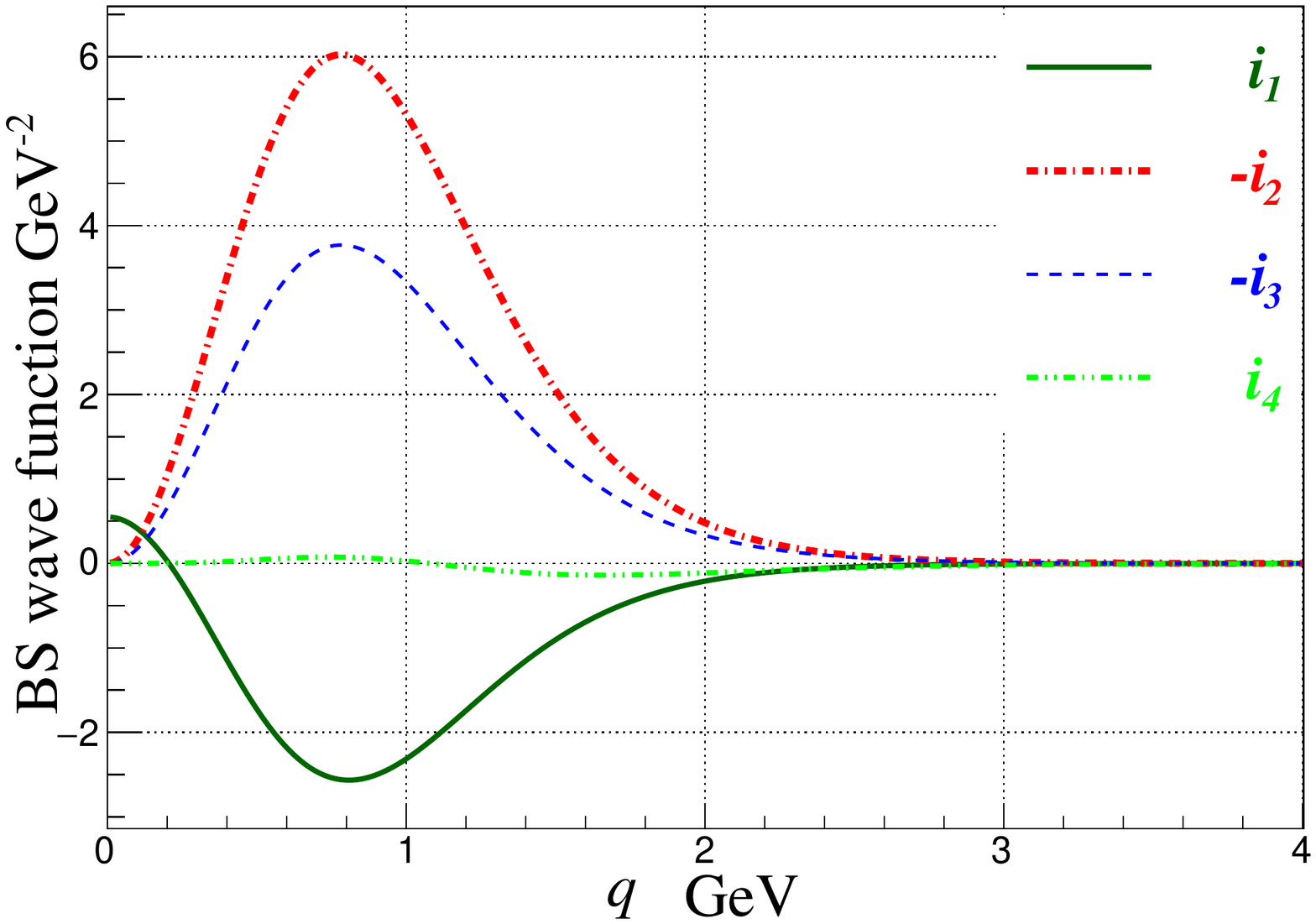} \label{Fig-wave-2-n4}}
\caption{Salpeter wave functions of the  $cc\bar c\bar c$ tetraquarks with $J^{PC}=0^{++}$, $1^{+-}$, and $2^{++}$.}\label{Fig-wave-cccc}
\vspace{0.5em}
\end{figure}

\autoref{Fig-wave-cccc} shows the obtained Salpeter wave functions for the first four $(cc\bar c\bar c)$ states obtained by the method I. Notice that the $D$ partial wave gives important contribution in the third and the forth states, and the possible $D$ or $G$-wave mixing are included naturally. The obtained wave functions show the rich information of the inner structure, and can be further used to do precise calculations on the decays, magnetic moments, or other properties of the tetraquarks.

\section{Summary}
In this work, we study the compact tetraquark states with fully heavy quark contents $QQ\bar Q\bar Q$ based on the Bethe-Salpeter equation. The compact tetraquark is taken as the bound state of the diquark-antidiquark where the (anti)diquark form factors are calculated with and without considering the confinement item respectively. {\color{black} In addition, a propagator-like form factor is also used to calculate the tetrequark states and the corresponding parameter dependence is also investigated.} Under the instantaneous approximation, the three-dimensional (Bethe-)Salpeter equation of the tetraquarks  are derived, and the Salpeter wave functions with $J^{PC}=0^{++}$, $1^{+-}$, and $2^{++}$ are then constructed and solved numerically to obtain the corresponding mass spectra of the tetraquarks. {\color{black} The compact tetraquarks as the bound states of the diquarks and antidiquarks follows a natural step of our previous work to deal with doubly heavy baryons. Therefore some results and approaches, such as the relevant interaction kernel, diquark constituent masses and form factors, the reduction of BSE, etc., can be directly adopted or extended to deal with compact tetraquarks, which allows us to take a systematic, relativistic, and unified framework to treat the heavy hadron systems, including the mesons, diquarks, baryons, and tetraquarks. Also notice all the parameters used to deal with the tetraquarks have already been fixed by the meson data and we do not introduce any free parameters. The well behaviors of the obtained wave functions allow to do the further precise calculations on the decays or other properties of the tetraquarks.} 

Our results show that the three ground states of $cc\bar c\bar c$ locate in the mass range $6.4\sim6.5\,\si{GeV}$, and the $bb\bar b\bar b$ states in mass range $19.2\sim19.3\,\si{GeV}$. Based on the obtained results, the LHCb's observation $X(6900)$ is less likely to be the ground states of compact $cc\bar c\bar c$ tetraquarks but might be the first or second radially excited states. The obtained relativistic wave functions are based on the good quantum numbers and can naturally include the mixing effects from the possible $D$(or $G$)-wave components.

\acknowledgments
This work is supported by the National Natural Science Foundation of China\,(NSFC) under Grant Nos.\,12005169, 12075073, 12075301, 11821505 and 11947302. It is also supported by the Natural Science Basic Research Program of Shaanxi\,(Program No.\,2021JQ-074), and the Fundamental Research Funds for the Central Universities.




\begin{thebibliography}{10}
\expandafter\ifx\csname url\endcsname\relax
  \def\url#1{\texttt{#1}}\fi
\expandafter\ifx\csname urlprefix\endcsname\relax\def\urlprefix{URL }\fi
\expandafter\ifx\csname href\endcsname\relax
  \def\href#1#2{#2} \def\path#1{#1}\fi

\bibitem{GellMann1964}
M.~Gell-Mann, Phys. Lett. 8 (1964) 214--215.
\newblock \href {https://doi.org/10.1016/S0031-9163(64)92001-3}
  {\path{DOI:10.1016/S0031-9163(64)92001-3}}.

\bibitem{Zweig1964}
G.~Zweig, Report No. CERN-TH-401 (1964).

\bibitem{LHCb2015-Pc}
R.~Aaij, et~al., Phys. Rev. Lett. 115 (2015) 072001.
\newblock \href {http://arxiv.org/abs/1507.03414} {\path{arXiv:1507.03414}},
  \href {https://doi.org/10.1103/PhysRevLett.115.072001}
  {\path{DOI:10.1103/PhysRevLett.115.072001}}.

\bibitem{LHCb2019-Pc}
R.~Aaij, et~al., Phys. Rev. Lett. 122~(22) (2019) 222001.
\newblock \href {http://arxiv.org/abs/1904.03947} {\path{arXiv:1904.03947}},
  \href {https://doi.org/10.1103/PhysRevLett.122.222001}
  {\path{DOI:10.1103/PhysRevLett.122.222001}}.

\bibitem{LHCb2020-X6900}
R.~Aaij, et~al., Sci. Bull. 65~(23) (2020) 1983--1993.
\newblock \href {http://arxiv.org/abs/2006.16957} {\path{arXiv:2006.16957}},
  \href {https://doi.org/10.1016/j.scib.2020.08.032}
  {\path{DOI:10.1016/j.scib.2020.08.032}}.

\bibitem{ZhaoJX2020}
J.~Zhao, S.~Shi, P.~Zhuang, Phys. Rev. D 102~(11) (2020) 114001.
\newblock \href {http://arxiv.org/abs/2009.10319} {\path{arXiv:2009.10319}},
  \href {https://doi.org/10.1103/PhysRevD.102.114001}
  {\path{DOI:10.1103/PhysRevD.102.114001}}.

\bibitem{Gordillo2020}
M.~C. Gordillo, F.~De~Soto, J.~Segovia, Phys. Rev. D 102~(11) (2020) 114007.
\newblock \href {http://arxiv.org/abs/2009.11889} {\path{arXiv:2009.11889}},
  \href {https://doi.org/10.1103/PhysRevD.102.114007}
  {\path{DOI:10.1103/PhysRevD.102.114007}}.

\bibitem{Mutuk2021}
H.~Mutuk, Eur. Phys. J. C 81~(4) (2021) 367.
\newblock \href {http://arxiv.org/abs/2104.11823} {\path{arXiv:2104.11823}},
  \href {https://doi.org/10.1140/epjc/s10052-021-09176-8}
  {\path{DOI:10.1140/epjc/s10052-021-09176-8}}.

\bibitem{Giron2020}
J.~F. Giron, R.~F. Lebed, Phys. Rev. D 102~(7) (2020) 074003.
\newblock \href {http://arxiv.org/abs/2008.01631} {\path{arXiv:2008.01631}},
  \href {https://doi.org/10.1103/PhysRevD.102.074003}
  {\path{DOI:10.1103/PhysRevD.102.074003}}.

\bibitem{Lundhammar2020}
P.~Lundhammar, T.~Ohlsson, Phys. Rev. D 102~(5) (2020) 054018.
\newblock \href {http://arxiv.org/abs/2006.09393} {\path{arXiv:2006.09393}},
  \href {https://doi.org/10.1103/PhysRevD.102.054018}
  {\path{DOI:10.1103/PhysRevD.102.054018}}.

\bibitem{DengCR2021}
C.~Deng, H.~Chen, J.~Ping, Phys. Rev. D 103~(1) (2021) 014001.
\newblock \href {http://arxiv.org/abs/2003.05154} {\path{arXiv:2003.05154}},
  \href {https://doi.org/10.1103/PhysRevD.103.014001}
  {\path{DOI:10.1103/PhysRevD.103.014001}}.

\bibitem{WengXZ2021}
X.-Z. Weng, X.-L. Chen, W.-Z. Deng, S.-L. Zhu, Phys. Rev. D 103~(3) (2021)
  034001.
\newblock \href {http://arxiv.org/abs/2010.05163} {\path{arXiv:2010.05163}},
  \href {https://doi.org/10.1103/PhysRevD.103.034001}
  {\path{DOI:10.1103/PhysRevD.103.034001}}.

\bibitem{WangZG2020}
Z.-G. Wang, Chin. Phys. C 44~(11) (2020) 113106.
\newblock \href {http://arxiv.org/abs/2006.13028} {\path{arXiv:2006.13028}},
  \href {https://doi.org/10.1088/1674-1137/abb080}
  {\path{DOI:10.1088/1674-1137/abb080}}.

\bibitem{LuQF2020}
Q.-F. L\"u, D.-Y. Chen, Y.-B. Dong, Eur. Phys. J. C 80~(9) (2020) 871.
\newblock \href {http://arxiv.org/abs/2006.14445} {\path{arXiv:2006.14445}},
  \href {https://doi.org/10.1140/epjc/s10052-020-08454-1}
  {\path{DOI:10.1140/epjc/s10052-020-08454-1}}.

\bibitem{Faustov2020}
R.~N. Faustov, V.~O. Galkin, E.~M. Savchenko, Phys. Rev. D 102~(11) (2020)
  114030.
\newblock \href {http://arxiv.org/abs/2009.13237} {\path{arXiv:2009.13237}},
  \href {https://doi.org/10.1103/PhysRevD.102.114030}
  {\path{DOI:10.1103/PhysRevD.102.114030}}.

\bibitem{Bedolla2020}
M.~A. Bedolla, J.~Ferretti, C.~D. Roberts, E.~Santopinto, Eur. Phys. J. C
  80~(11) (2020) 1004.
\newblock \href {http://arxiv.org/abs/1911.00960} {\path{arXiv:1911.00960}},
  \href {https://doi.org/10.1140/epjc/s10052-020-08579-3}
  {\path{DOI:10.1140/epjc/s10052-020-08579-3}}.

\bibitem{KeHW2021}
H.-W. Ke, X.~Han, X.-H. Liu, Y.-L. Shi, Eur. Phys. J. C 81~(5) (2021) 427.
\newblock \href {http://arxiv.org/abs/2103.13140} {\path{arXiv:2103.13140}},
  \href {https://doi.org/10.1140/epjc/s10052-021-09229-y}
  {\path{DOI:10.1140/epjc/s10052-021-09229-y}}.

\bibitem{ZhuRL2021}
R.~Zhu, Nucl. Phys. B 966 (2021) 115393.
\newblock \href {http://arxiv.org/abs/2010.09082} {\path{arXiv:2010.09082}},
  \href {https://doi.org/10.1016/j.nuclphysb.2021.115393}
  {\path{DOI:10.1016/j.nuclphysb.2021.115393}}.

\bibitem{ChaoKT2020}
K.-T. Chao, S.-L. Zhu, Sci. Bull. 65~(23) (2020) 1952--1953.
\newblock \href {http://arxiv.org/abs/2008.07670} {\path{arXiv:2008.07670}},
  \href {https://doi.org/10.1016/j.scib.2020.08.031}
  {\path{DOI:10.1016/j.scib.2020.08.031}}.

\bibitem{Karliner2020}
M.~Karliner, J.~L. Rosner, Phys. Rev. D 102~(11) (2020) 114039.
\newblock \href {http://arxiv.org/abs/2009.04429} {\path{arXiv:2009.04429}},
  \href {https://doi.org/10.1103/PhysRevD.102.114039}
  {\path{DOI:10.1103/PhysRevD.102.114039}}.

\bibitem{GuoZH2021}
Z.-H. Guo, J.~A. Oller, Phys. Rev. D 103~(3) (2021) 034024.
\newblock \href {http://arxiv.org/abs/2011.00978} {\path{arXiv:2011.00978}},
  \href {https://doi.org/10.1103/PhysRevD.103.034024}
  {\path{DOI:10.1103/PhysRevD.103.034024}}.

\bibitem{DongXK2021}
X.-K. Dong, V.~Baru, F.-K. Guo, C.~Hanhart, A.~Nefediev, Phys. Rev. Lett.
  126~(13) (2021) 132001.
\newblock \href {http://arxiv.org/abs/2009.07795} {\path{arXiv:2009.07795}},
  \href {https://doi.org/10.1103/PhysRevLett.126.132001}
  {\path{DOI:10.1103/PhysRevLett.126.132001}}.

\bibitem{WangJZ2021}
J.-Z. Wang, D.-Y. Chen, X.~Liu, T.~Matsuki, Phys. Rev. D 103~(7) (2021)
  L071503.
\newblock \href {http://arxiv.org/abs/2008.07430} {\path{arXiv:2008.07430}},
  \href {https://doi.org/10.1103/PhysRevD.103.L071503}
  {\path{DOI:10.1103/PhysRevD.103.L071503}}.

\bibitem{WanBD2020}
B.-D. Wan, C.-F. Qiao (12 2020).
\newblock \href {http://arxiv.org/abs/2012.00454} {\path{arXiv:2012.00454}}.

\bibitem{ZhuJW2020}
J.-W. Zhu, X.-D. Guo, R.-Y. Zhang, W.-G. Ma, X.-Q. Li (11 2020).
\newblock \href {http://arxiv.org/abs/2011.07799} {\path{arXiv:2011.07799}}.

\bibitem{LiQ2020}
Q.~Li, C.-H. Chang, S.-X. Qin, G.-L. Wang, Chin. Phys. C 44 (2020) 013102.
\newblock \href {http://arxiv.org/abs/1903.02282} {\path{arXiv:1903.02282}},
  \href {https://doi.org/10.1088/1674-1137/44/1/013102}
  {\path{DOI:10.1088/1674-1137/44/1/013102}}.

\bibitem{Brown2014}
Z.~S. Brown, W.~Detmold, S.~Meinel, K.~Orginos, Phys. Rev. D 90~(9) (2014)
  094507.
\newblock \href {http://arxiv.org/abs/1409.0497} {\path{arXiv:1409.0497}},
  \href {https://doi.org/10.1103/PhysRevD.90.094507}
  {\path{DOI:10.1103/PhysRevD.90.094507}}.

\bibitem{SB1951}
E.~E. Salpeter, H.~A. Bethe, Phys. Rev. 84 (1951) 1232--1242.
\newblock \href {https://doi.org/10.1103/PhysRev.84.1232}
  {\path{DOI:10.1103/PhysRev.84.1232}}.

\bibitem{Salpeter1952}
E.~E. Salpeter, Phys. Rev. 87 (1952) 328--343.
\newblock \href {https://doi.org/10.1103/PhysRev.87.328}
  {\path{DOI:10.1103/PhysRev.87.328}}.

\bibitem{Chang2005A}
C.-H. Chang, J.-K. Chen, X.-Q. Li, G.-L. Wang, Commun. Theor. Phys. 43 (2005)
  113--118.
\newblock \href {http://arxiv.org/abs/hep-ph/0406050}
  {\path{arXiv:hep-ph/0406050}}, \href
  {https://doi.org/10.1088/0253-6102/43/1/023}
  {\path{DOI:10.1088/0253-6102/43/1/023}}.

\bibitem{Chang2010}
C.~H. Chang, G.~L. Wang, Sci. China Phys. Mech. Astron. 53 (2010) 2005--2018.
\newblock \href {http://arxiv.org/abs/1003.3827} {\path{arXiv:1003.3827}},
  \href {https://doi.org/10.1007/s11433-010-4156-1}
  {\path{DOI:10.1007/s11433-010-4156-1}}.

\bibitem{LiQ2019A}
Q.~Li, T.~Wang, Y.~Jiang, G.-L. Wang, C.-H. Chang, Phys. Rev. D 100~(7) (2019)
  076020.
\newblock \href {http://arxiv.org/abs/1802.06351} {\path{arXiv:1802.06351}},
  \href {https://doi.org/10.1103/PhysRevD.100.076020}
  {\path{DOI:10.1103/PhysRevD.100.076020}}.

\bibitem{XuH2020}
H.~Xu, Q.~Li, C.-H. Chang, G.-L. Wang, Phys. Rev. D 101~(5) (2020) 054037.
\newblock \href {http://arxiv.org/abs/2001.02980} {\path{arXiv:2001.02980}},
  \href {https://doi.org/10.1103/PhysRevD.101.054037}
  {\path{DOI:10.1103/PhysRevD.101.054037}}.

\bibitem{Chang2005}
C.-H. Chang, C.~Kim, G.-L. Wang, Phys. Lett. B 623 (2005) 218--226.
\newblock \href {https://doi.org/10.1016/j.physletb.2005.07.059}
  {\path{DOI:10.1016/j.physletb.2005.07.059}}.

\bibitem{WangZ2012A}
Z.-H. Wang, G.-L. Wang, C.-H. Chang, J. Phys. G: Nucl. Part. Phys. 39 (2012)
  015009.
\newblock \href {http://arxiv.org/abs/1107.0474} {\path{arXiv:1107.0474}},
  \href {https://doi.org/10.1088/0954-3899/39/1/015009}
  {\path{DOI:10.1088/0954-3899/39/1/015009}}.

\bibitem{WangT2013}
T.~Wang, G.-L. Wang, H.-F. Fu, W.-L. Ju, JHEP 07 (2013) 120.
\newblock \href {http://arxiv.org/abs/1305.1067} {\path{arXiv:1305.1067}},
  \href {https://doi.org/10.1007/JHEP07(2013)120}
  {\path{DOI:10.1007/JHEP07(2013)120}}.

\bibitem{WangT2013A}
T.~Wang, G.-L. Wang, W.-L. Ju, Y.~Jiang, JHEP 03 (2013) 110.
\newblock \href {http://arxiv.org/abs/1303.1563} {\path{arXiv:1303.1563}},
  \href {https://doi.org/10.1007/JHEP03(2013)110}
  {\path{DOI:10.1007/JHEP03(2013)110}}.

\bibitem{LiQ2016}
Q.~Li, T.~Wang, Y.~Jiang, H.~Yuan, G.-L. Wang, Eur. Phys. J. C 76~(8) (2016)
  454.
\newblock \href {https://doi.org/10.1140/epjc/s10052-016-4306-3}
  {\path{DOI:10.1140/epjc/s10052-016-4306-3}}.

\bibitem{LiQ2017}
Q.~Li, T.~Wang, Y.~Jiang, H.~Yuan, T.~Zhou, G.-L. Wang, Eur. Phys. J. C 77~(1)
  (2017) 12.
\newblock \href {https://doi.org/10.1140/epjc/s10052-016-4588-5}
  {\path{DOI:10.1140/epjc/s10052-016-4588-5}}.

\bibitem{LiQ2017A}
Q.~Li, Y.~Jiang, T.~Wang, H.~Yuan, G.-L. Wang, C.-H. Chang, Eur. Phys. J. C
  77~(5) (2017) 297.
\newblock \href {http://arxiv.org/abs/1701.03252} {\path{arXiv:1701.03252}},
  \href {https://doi.org/10.1140/epjc/s10052-017-4865-y}
  {\path{DOI:10.1140/epjc/s10052-017-4865-y}}.

\bibitem{Chao1992}
K.-T. Chao, Y.-B. Ding, D.-H. Qin, Commun. Theor. Phys. 18 (1992) 321--326.

\bibitem{DingYB1993}
Y.-B. Ding, K.-T. Chao, D.-H. Qin, Chin. Phys. Lett. 10 (1993) 460--463.
\newblock \href {https://doi.org/10.1088/0256-307X/10/8/004}
  {\path{DOI:10.1088/0256-307X/10/8/004}}.

\bibitem{DingYB1995}
Y.-B. Ding, K.-T. Chao, D.-H. Qin, Phys. Rev. D 51 (1995) 5064--5068.
\newblock \href {http://arxiv.org/abs/hep-ph/9502409}
  {\path{arXiv:hep-ph/9502409}}, \href
  {https://doi.org/10.1103/PhysRevD.51.5064}
  {\path{DOI:10.1103/PhysRevD.51.5064}}.

\bibitem{Kim2004}
C.~S. Kim, G.-L. Wang, Phys. Lett. B 584 (2004) 285--293.
\newblock \href {http://arxiv.org/abs/hep-ph/0309162}
  {\path{arXiv:hep-ph/0309162}}, \href
  {https://doi.org/10.1016/j.physletb.2004.01.058}
  {\path{DOI:10.1016/j.physletb.2004.01.058}}.

\bibitem{Eichten1978}
E.~Eichten, K.~Gottfried, T.~Kinoshita, K.~D. Lane, T.-M. Yan, Phys. Rev. D 17
  (1978) 3090.
\newblock \href {https://doi.org/10.1103/PhysRevD.17.3090,
  10.1103/physrevd.21.313.2} {\path{DOI:10.1103/PhysRevD.17.3090,
  10.1103/physrevd.21.313.2}}.

\bibitem{Eichten1980}
E.~Eichten, K.~Gottfried, T.~Kinoshita, K.~D. Lane, T.-M. Yan, Phys. Rev. D 21
  (1980) 203.
\newblock \href {https://doi.org/10.1103/PhysRevD.21.203}
  {\path{DOI:10.1103/PhysRevD.21.203}}.

\bibitem{Laermann1986}
E.~Laermann, F.~Langhammer, I.~Schmitt, P.~M. Zerwas, Phys. Lett. B 173 (1986)
  437--442.
\newblock \href {https://doi.org/10.1016/0370-2693(86)90411-9}
  {\path{DOI:10.1016/0370-2693(86)90411-9}}.

\bibitem{Born1989}
K.~D. Born, E.~Laermann, N.~Pirch, T.~F. Walsh, P.~M. Zerwas, Phys. Rev. D 40
  (1989) 1653--1663.
\newblock \href {https://doi.org/10.1103/PhysRevD.40.1653}
  {\path{DOI:10.1103/PhysRevD.40.1653}}.

\bibitem{Cahill1987}
R.~T. Cahill, C.~D. Roberts, J.~Praschifka, Phys. Rev. D 36 (1987) 2804.
\newblock \href {https://doi.org/10.1103/PhysRevD.36.2804}
  {\path{DOI:10.1103/PhysRevD.36.2804}}.

\bibitem{Maris2002}
P.~Maris, Few Body Syst. 32 (2002) 41--52.
\newblock \href {http://arxiv.org/abs/nucl-th/0204020}
  {\path{arXiv:nucl-th/0204020}}, \href
  {https://doi.org/10.1007/s00601-002-0111-7}
  {\path{DOI:10.1007/s00601-002-0111-7}}.

\bibitem{Roberts2011}
H.~L.~L. Roberts, L.~Chang, I.~C. Cloet, C.~D. Roberts, Few Body Syst. 51
  (2011) 1--25.
\newblock \href {http://arxiv.org/abs/1101.4244} {\path{arXiv:1101.4244}},
  \href {https://doi.org/10.1007/s00601-011-0225-x}
  {\path{DOI:10.1007/s00601-011-0225-x}}.

\bibitem{LiuMS2019}
M.-S. Liu, Q.-F. L\"u, X.-H. Zhong, Q.~Zhao, Phys. Rev. D 100~(1) (2019)
  016006.
\newblock \href {http://arxiv.org/abs/1901.02564} {\path{arXiv:1901.02564}},
  \href {https://doi.org/10.1103/PhysRevD.100.016006}
  {\path{DOI:10.1103/PhysRevD.100.016006}}.

\bibitem{WangGJ2019}
G.-J. Wang, L.~Meng, S.-L. Zhu, Phys. Rev. D 100~(9) (2019) 096013.
\newblock \href {http://arxiv.org/abs/1907.05177} {\path{arXiv:1907.05177}},
  \href {https://doi.org/10.1103/PhysRevD.100.096013}
  {\path{DOI:10.1103/PhysRevD.100.096013}}.

\bibitem{ChenW2017}
W.~Chen, H.-X. Chen, X.~Liu, T.~G. Steele, S.-L. Zhu, Phys. Lett. B 773 (2017)
  247--251.
\newblock \href {http://arxiv.org/abs/1605.01647} {\path{arXiv:1605.01647}},
  \href {https://doi.org/10.1016/j.physletb.2017.08.034}
  {\path{DOI:10.1016/j.physletb.2017.08.034}}.

\bibitem{PDG2014}
K.~A. Olive, et~al., Chin. Phys. C 38 (2014) 090001.
\newblock \href {https://doi.org/10.1088/1674-1137/38/9/090001}
  {\path{DOI:10.1088/1674-1137/38/9/090001}}.

\end{thebibliography}

\end{document}